\documentclass[12pt]{iopart}
\usepackage{epsfig}
\usepackage{graphicx}

\newcommand{\pprime}{{\prime\prime}}

\newcommand{\bra}{\langle}
\newcommand{\ket}{\rangle}

\newcommand{\order}{{\mathcal O}}

\newcommand{\be}{\begin{equation}}
\newcommand{\ee}{\end{equation}}
\newcommand{\bd}{\begin{displaymath}}
\newcommand{\ed}{\end{displaymath}}

\newcommand{\R}{{\rm I\!R}}

\newcommand{\bc}{\ensuremath{\mathbf{c}}}

\newcommand{\bes}{\ensuremath{\mathbf{s}}}

\newcommand{\btau}{{\mbox{\boldmath $\tau$}}}

\newcommand{\bsigma}{{\mbox{\boldmath $\sigma$}}}

\newcommand{\here}{\makebox(0,0)}

\newcommand{\atanh}{{\rm atanh}}

\begin{document}\small

\title{Spin models on random graphs with controlled topologies beyond degree constraints}

\author{CJ P\'erez Vicente$^\dag$ and ACC Coolen$^{\S\diamond}$}
\address{\dag ~Departament de F\'isica Fonamental, Facultat de F\'isica,
 Universitat de Barcelona, 08028 Barcelona, Spain}
\address{\S ~ Department of Mathematics, King's College London, The Strand,
London WC2R 2LS, United Kingdom}
\address{$\diamond$ Randall Division of Cell and Molecular Biophysics,
King's College London, New Hunt's House, London SE1 1UL, United Kingdom}

\begin{abstract}
We study Ising spin models on finitely connected random interaction graphs
which are drawn from an ensemble in which not only the degree
distribution $p(k)$ can be chosen arbitrarily, but which allows for further fine-tuning of the topology
via preferential attachment of edges on the basis of an arbitrary function $Q(k,k^\prime)$ of the degrees of the
vertices involved. We solve these models using finite connectivity equilibrium replica theory, within the replica symmetric ansatz.
In our ensemble of graphs, phase diagrams of the spin system are found to depend no longer
only on the chosen degree distribution, but also on the choice made for $Q(k,k^\prime)$.
The increased ability to control interaction topology in solvable models beyond prescribing only the degree distribution
of the interaction graph enables a more accurate modeling of
real-world interacting particle systems by spin systems on suitably defined random graphs.
\end{abstract}

\pacs{75.10.Nr, 05.20.-y, 64.60.Cn}
\ead{\tt conrad@ffn.ub.es, ton.coolen@kcl.ac.uk}


\section{Introduction}

The study of spin systems on finitely connected random graphs started nearly thirty years ago
\cite{viana-bray85,kanter-sompo87,mezard-parisi87,mottishaw-dedominicis87,wong-sherrington88},
but has in the last decade enjoyed renewed popularity as a result of many successful
multi-disciplinary applications of the mathematical tools that it generates. The reason is clear: interacting particle systems
in the real world do not have full connectivity, but generally involve an average number of interaction partners per unit that is indeed limited.
Moreover, apart from carefully prepared pure samples of magnetic materials or crystalline solids,  the graph that represents which elements interact with each other appears in most
disciplines random in first approximation,
often with features that are remarkably universal across fields, such as a power law distribution of node degrees.
The mathematical
techniques used to study such systems are consequently being refined and applied in areas as diverse as spin-glass and glass modeling
\cite{monasson98,mezard-parisi01,mezard-parisi03,chiral,WNH,KuehnMourik},
 error correcting codes
\cite{murayama00,nakamura,nishimori,nikos,KS}, theoretical computer
science
\cite{kirkpatrick-selman94,monasson-zecchina98,monasson-zecchina982,monasson-zecchina99,VCising1,ColoringRG1,ColoringRG2},
recurrent neural networks
\cite{wemmenhove-coolen03,perez-skantzos03,arbitrary_degrees},
`small-world' networks \cite{SmallWorld,replTransMat,SmallWorldCavity}, socio-economic modeling \cite{IPC1,IPC2},
and CDMA communication \cite{CDMA1,CDMA2,CDMA3,CDMA4}.
Following a first wave of equilibrium studies, based on the finite connectivity replica method,
we are now also beginning to acquire mathematical techniques with which to bring
the dynamics of spin systems on finitely connected random graphs under control
 \cite{diluteDyn,StochDyn,ApprSch,ParDyn,DRTfc,Goos,LLmodel,Mozeika}.

In modeling real-world systems of the above type one places the dynamical variables (e.g. spins) on the vertices
of a graph, with edges connecting the variables that interact. The behaviour of the model depends strongly on the
choice made for the graph; however, this graph has to be drawn randomly to make analytical progress using disordered systems theory.
So one  tries to define a random graph ensemble with topological characteristics
that are as close as possible to those observed in the system to be modeled, while keeping the model solvable.
The state of the art in this respect is defining a random graph ensemble where edges are drawn
subject to the constraint that all vertex degrees are prescribed. Here the only information on the topology of the system
that is effectively carried over from the real world (apart from irrelevant site permutations) is the graph's degree distribution.
Since for most degree distributions there is still a large and diverse set of compatible microscopic graph realizations, with possibly
distinct macroscopic phenomenology for  spin dynamics and statics, one would like to increase the amount of topological information
embedded in the random graph ensemble, beyond prescribing just the degree distribution.

In this paper we present a simple class of Ising spin models on finitely connected random graphs, where
these graphs are drawn from ensembles in which not only the degree
distributions can be chosen arbitrarily and imposed as a constraint, but where in addition the edges are drawn in a way
that allows for preferential attachment on the basis of an  arbitrary function of the degrees of the two
vertices concerned (similar in spirit to the so-called hidden variable ensembles \cite{hv1,hv2,hv3,hv4,Boguna}).
The graphs thus generated are no longer characterized by their degree distribution alone, yet the associated spin models can still
be solved in thermal equilibrium using conventional finite connectivity replica techniques.
The graphs in the proposed ensemble remain effectively tree-like, and the clustering coefficients of randomly selected
vertices are zero with probability one (as in the standard finite connectivity ensembles with prescribed degrees only).
We solve our models with the replica symmetry (RS) ansatz,
and show which features of the phase diagrams and the dependencies
of observables on control parameters are identical across all graphs with the same degree distribution, and  which features
depend more specifically on the extra topological information that is included.

\section{Definitions}

We study a finitely connected and bond-disordered system of $N$
interacting Ising spins $\sigma_i\in\{-1,1\}$, in thermal
equilibrium characterized by the following Hamiltonian
\be
H(\bsigma)=-\sum_{i<j}c_{ij}\sigma_i J_{ij}\sigma_j~~~~~~~~
\bsigma=(\sigma_1,\ldots,\sigma_N) \label{eq:Hamiltonian}
\ee
  The frozen variables $c_{ij}\in\{0,1\}$ define the
connectivity of the system; they define a random graph, with vertices labeled
 by $i,j\in\{1,\ldots,N\}$ and with $c_{ij}=1$ if an only if $i$ and $j$ are connected by a link.
  We define $c_{ij}=c_{ji}$ and $c_{ii}=0$ for all $(i,j)$, and abbreviate $\bc=\{c_{ij}\}$. The bonds $J_{ij}\in\R$
 are drawn
randomly and independently from some distribution  $P(J)$. To characterize the topology of a graph $\bc$
we define for each vertex $i$ the degree
$k_i(\bc)=\sum_{j}c_{ij}$ (the number of links to this vertex), and the degree
distribution $p(k|\bc)=N^{-1}\sum_i \delta_{k,k_i(\bc)}$.
Thus the average connectivity of a graph $\bc$ is $\bra k\ket=\sum_{k\geq 0} kp(k|\bc)$, which we choose to be finite, even in the limit
$N\to\infty$.
We draw the graph $\bc$ randomly from an ensemble defined by a probability distribution ${\rm Prob}(\bc)$ in which
not only the degrees are constrained to take prescribed values $\{k_1,\ldots,k_N\}$,
but where the link
probabilities are modified further according to some function $Q(.,.)$ of the degrees of the two vertices
involved:
\begin{eqnarray}
\hspace*{-10mm}&&
{\rm Prob}(\bc)= \frac{1}{{\mathcal Z}_{N}}\prod_{i<j}\left[\frac{\bra k\ket
}{N}Q(k_i,k_j)\delta_{c_{ij},1}+\Big(1\!-\!\frac{\bra k\ket
}{N}Q(k_i,k_j)\Big)\delta_{c_{ij},0}\right]\prod_i\delta_{k_i,k_i(\bc)}
\label{eq:newer_connectivity}
 \\
 \hspace*{-10mm}&&
 {\mathcal Z}_{N}
=
\sum_{\bc}\prod_{i<j}\left[\frac{\bra k\ket}{N}Q(k_i,k_j)\delta_{c_{ij},1}+\Big(1\!-\!\frac{\bra k\ket}{N}Q(k_i,k_j)\Big)
\delta_{c_{ij},0}\right]\prod_i\delta_{k_i,k_i(\bc)}
 \end{eqnarray}
 The $N$ degrees are, in turn, drawn randomly from a prescribed distribution $p(k)$. Clearly we require $Q(k,k^\prime)\geq 0$ for all $k,k^\prime$.
 In order to ensure furthermore that for large $N$ such graphs can actually be found, we need to choose the function $Q(.,.)$ such that in formula (\ref{eq:newer_connectivity}) the partial
 measure $\prod_{i<j}[\ldots]$  is consistent with the average connectivity $\bra k\ket=\sum_k kp(k)$ imposed
 by the constraining factor $\prod_i\delta_{k_i,k_i(\bc)}$. Upon defining $Q(k,k^\prime)=Q(k^\prime,k)$, this is achieved when $\lim_{N\to\infty}N^{-2}\sum_{i\neq j}Q(k_i,k_j)=1$, so
  the function $Q(.,.)$ is to be chosen subject to
 \be
 Q(k,k^\prime)\geq 0~~\forall k,k^\prime~~~~~{\rm and}~~~~~\sum_{k,k^\prime\geq 0}p(k)p(k^\prime)Q(k,k^\prime)=1
 \label{eq:Qnormalization}
 \ee
  We note that an alternative but mathematically equivalent way to write the graph probabilities is obtained by applying to (\ref{eq:newer_connectivity}) the general identity
 \begin{eqnarray}
\prod_{i<j}[ A_{ij}\delta_{c_{ij},1}+B_{ij}\delta_{c_{ij},0}]&=& \Big(\prod_{i<j}B_{ij}\Big)\rme^{\sum_{i<j}c_{ij}[\log A_{ij}-\log B_{ij}]}
\end{eqnarray}
 which gives
 \begin{eqnarray}
{\rm Prob}(\bc)&=& \frac{1}{{\mathcal Z}_{N}}~\rme^{\sum_{i<j}c_{ij}\big\{\log [\bra k\ket
Q(k_i,k_j)/N]-\log[ 1\!-\!\bra k\ket
Q(k_i,k_j)/N]\big\}}
\prod_i\delta_{k_i,k_i(\bc)}
\label{eq:newer_connectivityB}
 \\
 {\mathcal Z}_{N}
&=&
\sum_{\bc}\rme^{\sum_{i<j}c_{ij}\big\{\log [\bra k\ket
Q(k_i,k_j)/N]-\log[ 1\!-\!\bra k\ket
Q(k_i,k_j)/N]\big\}}\prod_i\delta_{k_i,k_i(\bc)}
 \end{eqnarray}

 Given the above definitions, our objective is to calculate for the system (\ref{eq:Hamiltonian}) the asymptotic disorder-averaged
 free energy per spin $\overline{f}$, in order to find the phase diagrams for spin systems defined on typical graphs in the ensemble (\ref{eq:newer_connectivity}):
  \be
 \overline{f}=-\lim_{N\to\infty}\frac{1}{\beta N}\overline{\log \sum_{\bsigma}\rme^{-\beta H(\bsigma)}}
 =-\lim_{N\to\infty}\lim_{n\to 0}\frac{1}{\beta nN}\log \overline{\Big[\sum_{\bsigma}\rme^{-\beta H(\bsigma)}\Big]^n}
 \label{eq:f}
 \ee
 in which $\overline{\cdots}$ denotes averaging over the disorder in the problem, viz. the randomly drawn graph $\bc$ with statistics
 (\ref{eq:newer_connectivity}) and   the random bonds $\{J_{ij}\}$. This calculation is done with the finite connectivity
 replica method, within the replica symmetric (RS) ansatz. We will be particularly interested in the dependence of the phase diagrams on the choice made for $Q(.,.)$.
In the absence of the degree constraints, the function $Q(.,.)$ would have controlled the bond probabilities fully, via $\bra c_{ij}\ket=Q(k_i,k_j)\bra k\ket/N$. Here, in contrast, its role is to {\em deform} the measure imposed by the degree constraints, biasing the probabilities in those cases where there exist multiple graphs with the same degree distribution.

A relevant question to be asked at the start is whether the ensemble
deformation induced by $Q(k,k^\prime)$ can have sufficient impact in the thermodynamic limit on macroscopic observables to justify the present calculation.
For instance, upon calculating for the ensemble (\ref{eq:newer_connectivity}) the joint distribution of degrees and clustering coefficients for $N\to\infty$ one finds
that, like ensembles with degree constraints only, the clustering coefficients are zero with probability one, see \ref{app:clustering}.
However, upon reflection it becomes clear that the proposed ensemble deformation will generally
affect the system's phase diagram. A quick way to see this is to
compare the two choices $Q(k,k^\prime)=1$ and $Q(k,k^\prime)=\delta_{kk^\prime}/\sum_{k^\pprime}p^2(k^\pprime)$.
In the first case we return  to the degree-constrained ensemble in \cite{arbitrary_degrees}.
In the second case our ensemble describes graphs that are each composed of disconnected regular sub-graphs, of sizes $p(k)N$ for each $k$ with $p(k)>0$.
The transitions away from
the paramagnetic state will now be those corresponding to the sub-graph with the largest degree allowed by $p(k)$; these will only coincide
with those of $Q(k,k^\prime)=1$ when $p(k)=\delta_{k,c}$, i.e. when the degree distribution itself is that of a regular graph.

\section{Equilibrium replica analysis}

\subsection{Derivation of saddle-point equations}

As usual, we calculate (\ref{eq:f}) upon writing the  Kronecker $\delta$s of the degree constraints in integral form, using
$\delta_{nm}=(2\pi)^{-1}\int_{-\pi}^{\pi}\!\rmd\omega~\rme^{\rmi\omega(n-m)}$. This gives, after some rearranging of summations, factorization over the disorder variables. We define the short-hands $\bsigma_i=(\sigma_i^1,\ldots,\sigma_i^n)$, so that
\begin{eqnarray}
\overline{f}&=&
\lim_{N\to\infty}\lim_{n\to 0}\frac{1}{\beta nN}\Big\{
\log {\mathcal Z}_N-\log
\sum_{\bsigma_1\ldots\bsigma_N}\int_{-\pi}^\pi\prod_i\Big[\frac{\rmd\omega_i}{2\pi}\rme^{\rmi\omega_i k_i}\Big]
\nonumber
\\
&&\hspace*{15mm} \times
\prod_{i<j}\Big(
1
+\frac{\bra k\ket}{N}Q(k_i,k_j)[\int\!\rmd J~P(J)\rme^{\beta J\bsigma_i\cdot\bsigma_j-\rmi(\omega_i+\omega_j)}\!
-1]\Big)\Big\}
\nonumber
\\
&=&
\lim_{N\to\infty}\lim_{n\to 0}\frac{1}{\beta nN}\Big\{
\log {\mathcal Z}_N-\log
\sum_{\bsigma_1\ldots\bsigma_N}\int_{-\pi}^\pi\prod_i\Big[\frac{\rmd\omega_i}{2\pi}\rme^{\rmi\omega_i k_i}\Big]
\nonumber
\\
&&\times\exp\Big[
\frac{\bra k\ket}{2N}\sum_{ij} Q(k_i,k_j)[\int\!\rmd J~P(J)\rme^{\beta J\bsigma_i\cdot\bsigma_j-\rmi(\omega_i+\omega_j)}\!
-1]+\order(N^{0})\Big]\Big\}
\end{eqnarray}
We proceed towards a steepest descent integration by introducing for $\bsigma\in\{-1,1\}^n$ and $k\in\{0,1,2,\ldots\}$
the functions $D(k,\bsigma|\{\bsigma_i,\omega_i\})=
N^{-1}\sum_i\delta_{k,k_i}\delta_{\bsigma,\bsigma_i}\rme^{-\rmi\omega_i}$. They are introduced via the substitution of integrals
over appropriate $\delta$-distributions, written in integral form, viz.
\begin{eqnarray}
1&=& \int\!\frac{\rmd D(k,\bsigma)\rmd \hat{D}(k,\bsigma)}{2\pi/N}\rme^{\rmi N\hat{D}(k,\bsigma)[D(k,\bsigma)-
D(k,\bsigma|\{\bsigma_i,\omega_i\})]}
\end{eqnarray}
Upon using also $N^{-1}\sum_{ij} Q(k_i,k_j)=N+\order(\sqrt{N})$ due to (\ref{eq:Qnormalization}), and the short hand
$\{\rmd D \rmd \hat{D}\}=\prod_{k,\bsigma}D(k,\bsigma)\rmd \hat{D}(k,\bsigma)$ we then obtain
\begin{eqnarray}
\overline{f}&=&
\lim_{N\to\infty}\lim_{n\to 0}\frac{1}{\beta nN}\Big\{
\log {\mathcal Z}_N\!-\!\log
\int\!\{\rmd D\rmd \hat{D}\}\rme^{\rmi N\sum_{k\bsigma}\hat{D}(k,\bsigma)D(k,\bsigma)-\frac{1}{2}N\bra k\ket
+\order(N^{1/2})}
\nonumber
\\
&&\times\exp\Big[
\frac{1}{2}\bra k\ket N \sum_{kk^\prime}Q(k,k^\prime)\sum_{\bsigma\bsigma^\prime}D(k,\bsigma)D(k^\prime\!,\bsigma^\prime)
\int\!\rmd J~P(J)\rme^{\beta J\bsigma\cdot\bsigma^\prime}\!
\Big]
\nonumber
\\
&&
\times \exp\Big[ N\sum_k p(k)\log
\sum_{\bsigma}\int_{-\pi}^\pi\!\frac{\rmd\omega}{2\pi}\rme^{\rmi\omega k-\rmi
\hat{D}(k,\bsigma)\rme^{-\rmi\omega} }
\Big]
\Big\}
\end{eqnarray}
We next define $z=\lim_{N\to\infty}N^{-1}\log {\mathcal Z}_N$ (anticipating this limit to exist),
which allows us to evaluate $\overline{f}$ by steepest descent and write
\begin{eqnarray}
\overline{f}&=&\lim_{n\to 0} \frac{1}{n}{\rm extr}_{\{D,\hat{D}\}} f_n[\{D,\hat{D}\}]
\label{eq:f_steepest_descent}
\\
f_n[\{D,\hat{D}\}]&=&
-\frac{1}{\beta}\left\{
\rmi \sum_{k\bsigma}\hat{D}(k,\bsigma)D(k,\bsigma)-\frac{1}{2}\bra k\ket
-z
\right.
\nonumber
\\
&&
\left.
+\frac{1}{2}\bra k\ket \sum_{kk^\prime}Q(k,k^\prime)\sum_{\bsigma\bsigma^\prime}D(k,\bsigma)D(k^\prime\!,\bsigma^\prime)
\int\!\rmd J~P(J)\rme^{\beta J\bsigma\cdot\bsigma^\prime}\!
\right.
\nonumber
\\
&&
\left. +\sum_k p(k)\log
\sum_{\bsigma}\int_{-\pi}^\pi\!\frac{\rmd\omega}{2\pi}\rme^{\rmi\omega k-\rmi
\hat{D}(k,\bsigma)\rme^{-\rmi\omega} }
\right\}
\label{eq:fD}
\end{eqnarray}
The extremization in (\ref{eq:f_steepest_descent}) with respect to $\{D,\hat{D}\}$ gives the following saddle-point equations:
\begin{eqnarray}
\hat{D}(k,\bsigma)&=&
\rmi\bra k\ket \sum_{k^\prime}Q(k,k^\prime)\sum_{\bsigma^\prime}D(k^\prime\!,\bsigma^\prime)
\int\!\rmd J~P(J)\rme^{\beta J\bsigma\cdot\bsigma^\prime}
\\
 D(k,\bsigma)
 &=&
 \frac{p(k)\int_{-\pi}^\pi\!\rmd\omega~\rme^{\rmi\omega (k-1)-\rmi
\hat{D}(k,\bsigma)\rme^{-\rmi\omega} } }
{\sum_{\bsigma^\prime}\int_{-\pi}^\pi\!\rmd\omega~\rme^{\rmi\omega k-\rmi
\hat{D}(k,\bsigma^\prime)\rme^{-\rmi\omega} }}
\end{eqnarray}
The second of these equations can be simplified using the identity
\begin{eqnarray}
\int_{-\pi}^\pi\!\rmd\omega~\rme^{\rmi\omega\ell-\rmi
\hat{D}(k,\bsigma)\rme^{-\rmi\omega} } &=& \left\{\begin{array}{lll}
2\pi[-\rmi\hat{D}(k,\bsigma)]^{\ell}/\ell! &~~{\rm if}~~& \ell\geq 0
\\
0 &~~{\rm if}~~& \ell<0
\end{array}\right.
\label{eq:identity}
\end{eqnarray}
So, if we also re-define $\hat{D}(k,\bsigma)=\rmi \bra k\ket F(k,\bsigma)$, we arrive at
\begin{eqnarray}
F(k,\bsigma)&=& \sum_{k^\prime}Q(k,k^\prime)\sum_{\bsigma^\prime}D(k^\prime\!,\bsigma^\prime)
\int\!\rmd J~P(J)\rme^{\beta J\bsigma\cdot\bsigma^\prime}
\label{eq:SP_F}
\\
 D(k,\bsigma)
 &=& \frac{p(k)k}{\bra k\ket}
 \frac{F^{k-1}(k,\bsigma)}
{\sum_{\bsigma^\prime}F^k(k,\bsigma^\prime)}
\label{eq:SP_D}
\end{eqnarray}

\subsection{Simplified expression for free energy per spin}

Formula (\ref{eq:f_steepest_descent}) for the disorder-averaged free energy per spin,
which is to be evaluated at the relevant solution of the saddle-point equations
(\ref{eq:SP_F},\ref{eq:SP_D}), still contains the term $z=\lim_{N\to\infty}N^{-1}\log {\mathcal Z}_N$, which measures the effective
number of graphs in our ensemble (\ref{eq:newer_connectivity}). Since $z$ is independent of $\beta$ we can use the identity $\lim_{\beta\to 0}(\beta \overline{f})=-\log 2$ to find it. With (\ref{eq:identity}) and (\ref{eq:SP_F},\ref{eq:SP_D}) we first write (\ref{eq:f_steepest_descent})
 as
\begin{eqnarray}
\overline{f}&=&
\lim_{n\to 0}\frac{1}{\beta n}\Big\{
z+\bra k\ket-\bra k\ket\log\bra k\ket
-\sum_k p(k)\log \Big[
 \frac{1}{k!} \sum_{\bsigma} F^k(k,\bsigma)\Big]
\Big\}
\label{eq:f_further}
\end{eqnarray}
(with $D$ and $F$ taken at the relevant saddle-point).
Working out the saddle-point equations
for $\beta\to 0$ shows that there the two order parameter functions
$\{D,F\}$ become independent of $\bsigma$, viz. $D(k,\bsigma)=2^{-n}D(k)$ and $F(k,\bsigma)=F(k)$, where the latter obey
\begin{eqnarray}
F(k)= \sum_{k^\prime}\frac{p(k^\prime)k^\prime}{\bra k\ket} Q(k,k^\prime) F^{-1}(k^\prime),~~~~~~~~
D(k)= \frac{p(k)k}{\bra k\ket} F^{-1}(k)
\label{eq:SP_beta0}
\end{eqnarray}
Insertion into the equation $\lim_{\beta\to 0}(\beta \overline{f})=-\log 2$, together with (\ref{eq:f_further}), then leads us
after some further simple manipulations to the following formula, where $F(k)$ is the solution of (\ref{eq:SP_beta0}):
\begin{eqnarray}
z&=& \bra k\ket\log \bra k\ket-\bra k\ket +
\sum_{k} p(k)\log \Big[
 \frac{1}{k!} F^k(k)\Big]
 \label{eq:formula_for_z}
\end{eqnarray}
and hence
\begin{eqnarray}
\overline{f}&=&
-\lim_{n\to 0}\frac{1}{\beta n}
\sum_k p(k)\log \Big[ \sum_{\bsigma} [F(k,\bsigma)/F(k)]^k\Big]
\label{eq:f_RSB}
\end{eqnarray}
We note that for the non-deformed graph ensemble with degree constraints and finite connectivity statistics only, viz. $Q(k,k^\prime)=1$
for all $(k,k^\prime)$, the solution of (\ref{eq:SP_beta0}) would be $F(k)=1$ and $D(k)=p(k)k/\bra k\ket$. If we now define $\pi(k)=
 \rme^{-\bra k\ket}\bra k\ket^k/k!$, i.e. Poissonnian degree probabilities with average degree $\bra k\ket$, we see that
\begin{eqnarray}
z&=& z_{\rm nd} +\sum_{k} p(k)k\log F(k)
\label{eq:z_eqn}
\\
z_{\rm nd}&=&
\sum_{k} p(k)\log \pi(k)=-H_p-D(p||\pi)
\label{eq:z_nd}
\end{eqnarray}
with the entropy $H_p=-\sum_k p(k)\log p(k)\geq 0$ of the degree distribution and the Kullback-Leibler distance
$D(p||\pi)=\sum_k p(k)\log[p(k)/\pi(k)]\geq 0$ between the actual degree distribution $p(k)$ and the Poissonnian $\pi(k)$.
Since $z$ measures the effective number of graphs that can be generated from our ensemble, and $z_{\rm nd}$ is its value in the absence of deformation, it follows from (\ref{eq:z_eqn})
that we can define a simple measure $\Sigma_{\rm def}$ of the graph
specificity increase  resulting from the introduction of the deformation defined by a
 function $Q(k,k^\prime)$ as follows
\begin{eqnarray}
\Sigma_{\rm def}&=& -\sum_{k} p(k)k\log F(k)
\end{eqnarray}

\subsection{Replica symmetric theory}

To take the required limit $n\to 0$ in our formulae we make the ergodic or replica-symmetric (RS) ansatz.
The replica order parameter $D(k,\bsigma)$ must now be invariant under all replica permutations, and therefore have the following form, with
$\int\!\rmd h~D(k,h)=\sum_{\bsigma}D(k,\bsigma)$:
\begin{eqnarray}
D(k,\bsigma)&=&\int\!\rmd h~ D(k,h)\frac{\rme^{\beta
h\sum_{\alpha}\sigma_\alpha}}{[2\cosh(\beta h)]^n}
\label{eq:RSansatz}
\end{eqnarray}
We work out the implication of this ansatz for the order parameter $F(k,\bsigma)$, using the
 identity
 $f(\sigma)=e^{A\sigma}B$, with $A=\frac{1}{2}\log[f(1)/f(-1)]$
 and $B=\sqrt{f(1)f(-1)}$ (which holds for $\sigma=\pm 1$), as well as the identity
 $\frac{1}{2}\log[\cosh(x+y)/\cosh(x-y)]=\atanh[\tanh(x)\tanh(y)]$. This results in
\begin{eqnarray}
\hspace*{-20mm}
F(k,\bsigma)&=& \sum_{k^\prime}Q(k,k^\prime)\int\!\rmd h^\prime \rmd J~ D(k^\prime\!,h^\prime)P(J)
\prod_\alpha \frac{\cosh(\beta[J\sigma_\alpha+h^\prime])}{\cosh(\beta h^\prime)}
\nonumber
\\
\hspace*{-20mm}
&=&
\sum_{k^\prime}\!Q(k,k^\prime)\int\!\rmd h^\prime \rmd J~ D(k^\prime\!,h^\prime)P(J)G_n(h^\prime\!,J)
\rme^{\frac{1}{2}(\sum_\alpha \sigma_\alpha) \log[\cosh(\beta[J+h^\prime])/\cosh(\beta[J-h^\prime])]}
\nonumber
\\
\hspace*{-20mm}
&=& \int\!\rmd h~F(k,h)~ \rme^{\beta h\sum_\alpha \sigma_\alpha}
\label{eq:RSansatzF}
\end{eqnarray}
with
\begin{eqnarray}
 G_n(h,J)&=&\big[\cosh(\beta [h\!+\!J])\cosh(\beta
[h\!-\!J])/\cosh^2(\beta h)\big]^{n/2}
\\[1mm]
F(k,h)&=& \sum_{k^\prime}\!Q(k,k^\prime)\int\!\rmd h^\prime \rmd J~ D(k^\prime\!,h^\prime)P(J)G_n(h^\prime\!,J)
\nonumber
\\[-1mm]
&&\hspace*{15mm}\times \delta\big[h-\beta^{-1}\atanh[\tanh(\beta J)\tanh(\beta h^\prime)]\big]
\label{eq:SPRS_V}
\end{eqnarray}
 We will now have two new saddle-point equations, written in terms of the RS kernels $F$ and $D$.
The first equation is (\ref{eq:SPRS_V}). The second follows upon inserting (\ref{eq:RSansatz},\ref{eq:RSansatzF}) into
(\ref{eq:SP_D}):
\begin{eqnarray}
\hspace*{-15mm}
 \int\!\rmd h~ D(k,h)\frac{\rme^{\beta
h\sum_{\alpha}\sigma_\alpha}}{[2\cosh(\beta h)]^n}
 &=& \frac{p(k)k}{\bra k\ket}
 \frac{\int\!\prod_{\ell\leq k-1}[\rmd h_\ell F(k,h_\ell)] \rme^{\beta\sum_\alpha \sigma_\alpha \sum_{\ell\leq k-1}h_\ell}}
{\int\!\prod_{\ell\leq k}[\rmd h_\ell F(k,h_\ell)] [2\cosh(\beta \sum_{\ell\leq k}h_\ell)]^n}
\nonumber
\\
\hspace*{-15mm}
&=&
\frac{p(k)k}{\bra k\ket}\int\!dh~\rme^{\beta
h\sum_{\alpha}\sigma_\alpha}
\nonumber
\\
\hspace*{-15mm}
&&\times
 \frac{\int\!\prod_{\ell\leq k-1}[\rmd h_\ell F(k,h_\ell)] \delta[h-\sum_{\ell\leq k-1}h_\ell]}
{\int\!\prod_{\ell\leq k}[\rmd h_\ell F(k,h_\ell)] [2\cosh(\beta \sum_{\ell\leq k}h_\ell)]^n}
\end{eqnarray}
From this result we can read off the second order parameter equation.
Upon taking the limit $n\to 0$ in the latter result and our first equation (\ref{eq:SPRS_V}), we arrive at
the transparent expressions
\begin{eqnarray}
\hspace*{-15mm}
F(k,h)&=& \sum_{k^\prime}\!Q(k,k^\prime)\int\!\rmd h^\prime \rmd J~ D(k^\prime\!,h^\prime)P(J)
\delta\big[h-\frac{1}{\beta}\atanh[\tanh(\beta J)\tanh(\beta h^\prime)]\big]
\label{eq:RS1}
\\
\hspace*{-15mm}
 D(k,h)
 &=&
\frac{p(k)k}{\bra k\ket}
 \frac{\int\!\prod_{\ell\leq k-1}[\rmd h_\ell F(k,h_\ell)] \delta[h-\sum_{\ell\leq k-1}h_\ell]}
{[\int\!\rmd h^\prime F(k,h^\prime)]^k }
\label{eq:RS2}
\end{eqnarray}
Upon defining finally $D(k)=\int\!\rmd h~D(k,h)$ and $F(k)=\int\!\rmd h F(k,h)$, we discover that these last two integrals are
exactly the quantities in (\ref{eq:SP_beta0}), since upon integrating over $h$ in both equations (\ref{eq:RS1},\ref{eq:RS2}) they
reduce precisely to (\ref{eq:SP_beta0}). This does not seem to be a trivial result, since (\ref{eq:SP_beta0})
was derived from the properties of the random graph ensemble alone. Given these relations one is prompted automatically to define
$D(k,h)=D(h|k)D(k)$ and $F(k,h)=F(h|k)F(k)$, where now $\int\!\rmd h~D(h|k)=\int\!\rmd h~F(h|k)=1$.
Our RS order parameter equations thereby take the form
\begin{eqnarray}
\hspace*{-20mm}
F(h|k)&=& \sum_{k^\prime}\frac{Q(k,k^\prime)p(k^\prime)k^\prime}{\bra k\ket F(k)F(k^\prime)}\int\!\rmd h^\prime \rmd J~ D(h^\prime|k^\prime)P(J)
\delta\big[h\!-\!\frac{1}{\beta}\atanh[\tanh(\beta J)\tanh(\beta h^\prime)]\big]
\label{eq:RS1b}
\\
\hspace*{-20mm}
&&
 D(h|k)
 =
\int\!\prod_{\ell<k}[\rmd h_\ell F(h_\ell|k)] \delta\big[h-\sum_{\ell<k}h_\ell\big]
\label{eq:RS2b}
\\
\hspace*{-20mm}
&&
F(k)= \bra k\ket^{-1}\sum_{k^\prime}p(k^\prime)k^\prime Q(k,k^\prime) F^{-1}(k^\prime)
\label{eq:RS3b}
\end{eqnarray}
where $D(k)$ subsequently follows from the second identity in (\ref{eq:SP_beta0}), and where (\ref{eq:RS3b}) automatically
ensures that the solutions of (\ref{eq:RS1b},\ref{eq:RS2b}) are normalized.
If we write the free energy (\ref{eq:f_RSB}) for our present RS solution
in terms of the quantities in (\ref{eq:RS1b},\ref{eq:RS2b},\ref{eq:RS3b}) and take the limit $n\to 0$,
using relations such as $\sum_k D(k)F(k)=1$,
we find in a similar manner the remarkably simple result
\begin{eqnarray}
\overline{f}_{\rm RS}&=&-\frac{1}{\beta}\log 2
-\frac{1}{\beta}  \sum_{k}p(k)
\int\!\prod_{\ell\leq k}[\rmd h_\ell F(h_\ell|k)]\log
\cosh(\beta \sum_{\ell\leq k}h_\ell)
\label{eq:f_RS}
\end{eqnarray}

\subsection{Physical observables}

To find formulae for observables like $m=\lim_{N\to\infty}N^{-1}\sum_i\overline{\bra \sigma_i\ket}$
and $q=\lim_{N\to\infty}N^{-1}\sum_i\overline{\bra \sigma_i\ket^2}$, it will be convenient to calculate for an $n$-replica system
where $\bsigma_i=(\sigma_i^1,\ldots,\sigma_i^n)$, i.e. before the
limit $n\to 0$, the expectation value $P(k,\bsigma)=\lim_{N\to\infty}N^{-1}\sum_i\overline{\bra \delta_{k,k_i}\delta_{\bsigma,\bsigma_i}\ket}$.
We can calculate $P(k,\bsigma)$ using steps similar to those taken in the evaluation of the free energy, using (\ref{eq:fD}):
\begin{eqnarray}
 P(k,\bsigma)&=&
 \lim_{N\to\infty}N^{-1}\sum_i\overline{\left[
 \frac{\sum_{\bsigma_1\ldots\bsigma_N}\delta_{k,k_i}\delta_{\bsigma,\bsigma_i}\rme^{-\beta\sum_\alpha H(\bsigma^\alpha)}}
 {\sum_{\bsigma_1\ldots\bsigma_N}\rme^{-\beta\sum_\alpha H(\bsigma^\alpha)}}\right]}
\nonumber
\\
&=& \lim_{n^\prime\to -n}
 \lim_{N\to\infty}\frac{1}{N}\sum_i\delta_{k,k_i}\overline{
 \sum_{\bsigma^1\ldots\bsigma^{n+n^\prime}}\delta_{(\sigma_1,\ldots,\sigma_n),(\sigma_i^1,\ldots,\sigma_i^n)}
 \rme^{-\beta\sum_{\alpha=1}^{n+n^\prime} H(\bsigma^\alpha)}}
 \nonumber
\\
&=& \lim_{n^\prime\to -n}
 \lim_{N\to\infty}\frac{1}{N}\sum_i\delta_{k,k_i}
\int\!\{\rmd D\rmd \hat{D}\}\rme^{-\beta Nf_{n+n^\prime}[\{D,\hat{D}\}]}
\nonumber
\\
&&\times
 \left[\frac{\sum_{\sigma_{n+1},\ldots,\sigma_{n+n^\prime}}\int\!\rmd\omega~\rme^{\rmi\omega k-\rmi\hat{D}(k,\sigma_1,\ldots,\sigma_{n+n^\prime})\rme^{-\rmi \omega}}}
{\sum_{\sigma_1^\prime,\ldots\sigma^\prime_{n+n^\prime}}\int\!\rmd\omega~\rme^{\rmi\omega k-\rmi\hat{D}(k,\sigma_1^\prime,\ldots,\sigma_{n+n^\prime}^\prime)\rme^{-\rmi \omega}}}
\right]
\nonumber
\\
&=& \lim_{n^\prime\to -n}p(k) \left[\frac{\sum_{\sigma_{n+1},\ldots,\sigma_{n+n^\prime}}\int\!\rmd\omega~\rme^{\rmi\omega k+\bra k\ket F(k,\sigma_1,\ldots,\sigma_{n+n^\prime})\rme^{-\rmi \omega}}}
{\sum_{\sigma_1^\prime,\ldots\sigma^\prime_{n+n^\prime}}\int\!\rmd\omega~\rme^{\rmi\omega k+\bra k\ket F(k,\sigma_1^\prime,\ldots,\sigma_{n+n^\prime}^\prime)\rme^{-\rmi \omega}}}
\right]
\end{eqnarray}
where in the last line we now have to take $F(k,\sigma_1,\ldots,\sigma_{n+n^\prime})$ at the saddle-point of $f_{n+n^\prime}[\ldots]$,
with $\hat{D}[\ldots]=\rmi\bra k\ket F[\ldots]$, and where we have used $\lim_{n\to 0}{\rm extr}_{\{D,\hat{D}\}}f_n[\{D,\hat{D}\}]=0$.
Once more we can carry out the $\omega$-integrations and find
\begin{eqnarray}
 P(k,\bsigma)&=&
 \lim_{n^\prime\to -n}p(k) \left[\frac{\sum_{\sigma_{n+1},\ldots,\sigma_{n+n^\prime}}F^k(k,\sigma_1,\ldots,\sigma_{n+n^\prime})}
{\sum_{\sigma_1^\prime,\ldots\sigma^\prime_{n+n^\prime}}F^k(k,\sigma_1^\prime,\ldots,\sigma_{n+n^\prime}^\prime)}
\right]
\end{eqnarray}
In replica symmetric states, where $F(k,\bsigma)=\int\!\rmd h~F(k)F(h|k)\rme^{\beta h\sum_\alpha\sigma_\alpha}$,  we can carry out the remaining spin summations, giving
\begin{eqnarray}
\hspace*{-15mm}
 P_{\rm RS}(k,\bsigma)&=&
 \lim_{n^\prime\to -n}p(k) \left[\frac{\int\!\prod_{\ell\leq k}[\rmd h_\ell F(h_\ell|k)] [2\cosh(\beta\sum_{\ell\leq k}h_\ell)]^{n^\prime}
 \rme^{\beta(\sum_{\ell\leq k}h_\ell)(\sum_{\alpha\leq n} \sigma_\alpha)}}
{\int\!\prod_{\ell\leq k}[\rmd h_\ell F(h_\ell|k)]
[2\cosh(\beta\sum_{\ell\leq k}h_\ell)]^{n+n^\prime}}
\right]
\nonumber
\\
\hspace*{-15mm}
 &=&
 p(k)
 \int\!\prod_{\ell\leq k}[\rmd h_\ell F(h_\ell|k)]
 \frac{\rme^{\beta(\sum_{\ell\leq k}h_\ell)(\sum_{\alpha\leq n} \sigma_\alpha)}}
 {[2\cosh(\beta\sum_{\ell\leq k}h_\ell)]^{n}}
 \nonumber
 \\
 \hspace*{-15mm}
 &=& p(k)\int\!\rmd h~W(h|k)\frac{\rme^{\beta h\sum_{\alpha\leq n} \sigma_\alpha}}
 {[2\cosh(\beta h)]^{n}}
\end{eqnarray}
with the degree-conditioned effective field distribution
\begin{eqnarray}
W(h|k)&=&
 \int\!\prod_{\ell\leq k}[\rmd h_\ell F(h_\ell|k)]~\delta\Big[h-
 \sum_{\ell\leq k}h_\ell\Big]
 \label{eq:effective_fields}
\end{eqnarray}
The order parameters $m$ and $q$ can now be written in their usual form
\begin{eqnarray}
&&m=\int\!\rmd h~W(h)\tanh(\beta h),~~~~~~~~q=\int\!\rmd h~W(h)\tanh^2(\beta h)
\\
&& W(h)=\sum_k p(k)W(h|k)
\end{eqnarray}
and the free energy formula (\ref{eq:f_RS}) becomes simply
\begin{eqnarray}
\overline{f}_{\rm RS}&=&-\frac{1}{\beta}\log 2
-\frac{1}{\beta}\int\!\rmd h~W(h)\log
\cosh(\beta h)
\end{eqnarray}

\subsection{Simple solutions for special cases}

The simplest limit is $\beta\to 0$, where we expect a paramagnetic state. Setting $\beta=0$ in our order parameter equations (\ref{eq:RS1b},\ref{eq:RS2b},\ref{eq:RS3b}) indeed gives
$F(h|k)=D(h|k)=\delta(h)$, and $\lim_{\beta\to 0}(\beta \overline{f})=-\log 2$ as well as $m=q=0$.
As always in such systems one notes that $F(h|k)=D(h|k)=\delta(h)$ is in fact always a solution of (\ref{eq:RS1b},\ref{eq:RS2b},\ref{eq:RS3b}),
at any temperature.

A less trivial special case is the choice
$Q(k,k^\prime)=1$ for all $(k,k^\prime)$ (non-deformed graph ensemble), where we should be able to connect the present theory
to earlier results in literature. Here one has $F(k)=1$ and $D(k)=p(k)k/\bra k\ket$, and $F(h|k)$ is no longer dependent on $k$.
We may thus simply write $F(h|k)=F(h)$ (not to be confused with $F(k)$), so that upon eliminating $D(h|k)$ from (\ref{eq:RS1b}) via (\ref{eq:RS2b}), we are left with the order parameter equation
\begin{eqnarray}
\hspace*{-20mm}
F(h)&=& \sum_{k}\frac{p(k)k}{\bra k\ket}\int\!\rmd J~P(J)\int\!\prod_{\ell<k}[\rmd h_\ell F(h_\ell)]
\delta\Big[h\!-\!\frac{1}{\beta}\atanh[\tanh(\beta J)\tanh(\beta \sum_{\ell<k}h_\ell)]\Big]
\label{eq:Fconnect}
\end{eqnarray}
To establish the connection with earlier results for the Ising system on graphs taken from the non-deformed ensemble
we define a new field distribution $\tilde{W}(h)$
\begin{eqnarray}
\tilde{W}(h)&=& \sum_{k}\frac{p(k)k}{\bra k\ket}\int\!\prod_{\ell<k}[\rmd h_\ell F(h_\ell)] \delta\Big[h-\sum_{\ell<k}h_\ell\Big]
\end{eqnarray}
According to (\ref{eq:effective_fields}), the latter cavity field distribution $\tilde{W}(h)$ is expressed in terms of
the degree-conditioned effective field distributions
$W(h|k)$ via $\tilde{W}(h)=\sum_k p(k) k\bra k\ket^{-1}W(h|k-1)$.
Equation (\ref{eq:Fconnect}) now tells us that
\begin{eqnarray}
F(h)&=&\int\!\rmd J\rmd h^\prime P(J)\tilde{W}(h^\prime)
\delta\Big[h\!-\!\frac{1}{\beta}\atanh[\tanh(\beta J)\tanh(\beta h^\prime)]\Big]
\end{eqnarray}
and hence we find the following RS order parameter equation in terms of $\tilde{W}(h)$, which we recognize from earlier studies on Ising systems
with random graph ensembles that are given prescribed degree distributions:
\begin{eqnarray}
\hspace*{-20mm}
\tilde{W}(h)&=& \sum_{k}\frac{p(k)k}{\bra k\ket}\int\!\prod_{\ell<k}[\rmd J_\ell\rmd h_\ell P(J_\ell)\tilde{W}(h_\ell)] \delta\Big[h-\frac{1}{\beta}\sum_{\ell<k}\atanh[\tanh(\beta J_\ell)\tanh(\beta h_\ell)]\Big]
\end{eqnarray}
Similarly we can write the free energy (\ref{eq:f_RS}) for non-deformed ensembles in terms of $\tilde{W}(h)$:
\begin{eqnarray}
\overline{f}_{\rm RS}&=&-\frac{1}{\beta}\log 2
-\frac{1}{\beta}  \sum_{k}p(k)
\int\!\prod_{\ell\leq k}[\rmd h_\ell \rmd J_\ell P(J_\ell)\tilde{W}(h_\ell)]
\nonumber
\\
&&\hspace*{15mm}\times
\log
\cosh\Big(\sum_{\ell\leq k}\atanh[\tanh(\beta J_\ell)\tanh(\beta h_\ell)]\Big)
\end{eqnarray}
We should emphasize that the two field distributions $W(h)=\sum_k p(k) W(h|k)$ and $\tilde{W}(h)=\sum_k p(k\!+\!1)(k\!+\!1)\bra k\ket^{-1}W(h|k)$ are generally different. The obvious exceptions are Poissonnian degree distributions, where $p(k\!+\!1)(k\!+\!1)\bra k\ket^{-1}=p(k)$,
and systems where $W(h|k)$ does not vary with the degree $k$.

\section{Continuous phase transitions}

\subsection{Bifurcations away from the paramagnetic state}

Continuous bifurcations away from the paramagnetic (P) state $F(h|k)=\delta(h)$ are found in the usual manner, by expansion in moments of $F(h|k)$.
We assume the existence of a small parameter $\epsilon$ with  $0<|\epsilon|\ll 1$ such that $\int\!\rmd h~h^\ell F(h|k)=\order(\epsilon^\ell)$.
Let us first define $\epsilon_k=\int\!\rmd h~h F(h|k)$.
Multiplication of  (\ref{eq:RS1b},\ref{eq:RS2b}) by $h$, followed by integration over $h$ gives the lowest nontrivial order:
\begin{eqnarray}
\hspace*{-15mm}
\epsilon_k &=& \frac{1}{\beta}\sum_{k^\prime}\frac{Q(k,k^\prime)p(k^\prime)k^\prime}{\bra k\ket F(k)F(k^\prime)}\int\!\rmd J~P(J)
\int\!\prod_{\ell<k^\prime}[\rmd h_\ell F(h_\ell|k^\prime)]
\atanh[\tanh(\beta J)\tanh(\beta \sum_{\ell<k^\prime}h_\ell)]
\nonumber
\\
\hspace*{-15mm}
&=& \int\!\rmd J~P(J)\tanh(\beta J)\sum_{k^\prime}\frac{Q(k,k^\prime)p(k^\prime)k^\prime(k^\prime\!\!-\!1)}{\bra k\ket F(k)F(k^\prime)}
~\epsilon_{k^\prime}+\order(\epsilon^2)
\end{eqnarray}
Thus a continuous transition to a ferromagnetic (F) state occurs when the matrix with entries $M_{kk^\prime}\int\!\rmd J~P(J)\tanh(\beta J)$ has en eigenvalue one, where $k,k^\prime\in\{0,1,2,\ldots\}$ and where
\begin{eqnarray}
M_{kk^\prime}&=&
 \frac{Q(k,k^\prime)p(k^\prime)k^\prime(k^\prime\!\!-\!1)}{\bra k\ket F(k)F(k^\prime)}
\label{eq:defineM}
\end{eqnarray}
In a ferromagnetic state one will have a nonzero magnetization $m=\beta\sum_k p(k)k\epsilon_k+\order(\epsilon^2)$.
Similarly we can check what happens if $\int\!\rmd h~h F(h|k)=0$ for all $k$, so $m=0$, and the first nontrivial order
to bifurcate is $\epsilon^2$. This corresponds to a transition from a paramagnetic to a spin-glass (SG) state.
Now we define $\epsilon_k=\int\!\rmd h ~h^2F(h|k)=\order(\epsilon^2)$.
Multiplication of  (\ref{eq:RS1b},\ref{eq:RS2b}) by $h^2$, followed by integration over $h$ now gives
\begin{eqnarray}
\hspace*{-15mm}
\epsilon_k&=& \int\!\rmd J~P(J)\tanh^2(\beta J)
\sum_{k^\prime}\frac{Q(k,k^\prime)p(k^\prime)k^\prime}{\bra k\ket F(k)F(k^\prime)}
\int\!\prod_{\ell<k^\prime}[\rmd h_\ell F(h_\ell|k^\prime)]\sum_{\ell,\ell^\prime<k^\prime}h_\ell h_\ell^\prime
+\order(\epsilon^3)
\nonumber
\\
\hspace*{-15mm}&=&
\int\!\rmd J~P(J)\tanh^2(\beta J)
\sum_{k^\prime}\frac{Q(k,k^\prime)p(k^\prime)k^\prime(k^\prime\!\!-\!1)}{\bra k\ket F(k)F(k^\prime)}~\epsilon_{k^\prime}
+\order(\epsilon^3)
\end{eqnarray}
(where we have used $\int\!\rmd h~hF(h|k)=0$ to eliminate terms with $\ell\neq \ell^\prime$).
Thus a continuous transition to a spin-glass (SG) state occurs when the matrix with entries $M_{kk^\prime}\int\!\rmd J~P(J)\tanh^2(\beta J)$ has an eigenvalue one. The previous P$\to$F transitions mark bifurcations away from $(m,q)=(0,0)$ to states with $(m\neq 0,q\neq 0)$,
and the P$\to$SG transitions mark bifurcations away from $(m,q)=(0,0)$ to states with $(m=0,q\neq 0)$:
\begin{eqnarray}
{\rm P}\to{\rm F:}
&~~~~\sum_{k^\prime}M_{kk^\prime}\epsilon_{k^\prime}=\Lambda_{\rm F}
\epsilon_k~~~~~~& \Lambda^{-1}_{\rm F}=\int\!\rmd J~P(J)\tanh(\beta J)
\\
{\rm P}\to{\rm SG:}
&~~~~\sum_{k^\prime}M_{kk^\prime}\epsilon_{k^\prime}=\Lambda_{\rm SG}\epsilon_k
~~~~~~& \Lambda^{-1}_{\rm SG}=\int\!\rmd J~P(J)\tanh^2(\beta J)
\end{eqnarray}
with the matrix elements $M_{kk^\prime}$ as given in (\ref{eq:defineM}) and with $F(k)$ to be solved from (\ref{eq:RS3b}):
\begin{eqnarray}
F(k)= \bra k\ket^{-1}\sum_{k^\prime}p(k^\prime)k^\prime Q(k,k^\prime) F^{-1}(k^\prime)
\label{eq:eqnforF}
\end{eqnarray}
The physical transition occurs at the largest eigenvalue of
the matrix $M$. It will be of the type P$\to$F if
$\int\!\rmd J~P(J)\tanh(\beta J)>\int\!\rmd J~P(J)\tanh^2(\beta J)$, and
otherwise it will be P$\to$SG. If for the bond distribution we
choose the binary form
$P(J)=\frac{1}{2}(1+\eta)\delta(J-J_0)+\frac{1}{2}(1-\eta)\delta(J+J_0)$,
we can already deduce that the triple point (where the
phases P, F and SG come together) is found along the line
$\eta=\tanh(\beta J_0)$, with the P$\to$F transition being the
physical one for $\eta>\tanh(\beta J_0)$ and  P$\to$SG
transition being the physical one for $\eta<\tanh(\beta J_0)$.
This will be true irrespective of the choice made for the ensemble deformation function $Q(k,k^\prime)$.

\subsection{Analysis of the eigenvalue problems}

We will now analyze the bifurcation eigenvalue problem for the matrix (\ref{eq:defineM}), to be solved in conjunction
 with (\ref{eq:eqnforF}), for a number of graph ensembles, which in this paper are characterized by an ensemble deformation
function $Q(k,k^\prime)$ and a degree distribution $p(k)$.
The deformation function must always obey $Q(k,k^\prime)\geq 0$, $Q(k,k^\prime)=Q(k^\prime,k)$, and $\sum_{kk^\prime}p(k)p(k^\prime)Q(k,k^\prime)=1$. Upon writing the largest eigenvalue of the matrix (\ref{eq:defineM}) as $\lambda_{\rm max}(Q,p)$,
the continuous bifurcations away from the paramagnetic state occur for
\begin{eqnarray}
{\rm P}\to{\rm F:}
&~~~~& 1=\lambda_{\max}(Q,p)\int\!\rmd J~P(J)\tanh(\beta J)
\label{eq:PtoF}
\\
{\rm P}\to{\rm SG:}
&~~~~& 1=\lambda_{\rm max}(Q,p)\int\!\rmd J~P(J)\tanh^2(\beta J)
\label{eq:PtoSG}
\end{eqnarray}

\subsubsection*{Type I: Separable deformation functions.}

The simplest family of deformation functions are of the separable form
$Q(k,k^\prime)=g(k)g(k^\prime)/\bra g\ket^2$, with $g(k)\geq 0$
for all $k$, and $\bra g\ket=\sum_k p(k)g(k)>0$. The special choice $g(k)=1$ gives the ensemble without
deformation. For this family it follows immediately from (\ref{eq:eqnforF}) that $F(k)=g(k)/\bra g\ket$,
and (\ref{eq:defineM}) reduces to
\begin{eqnarray}
M_{kk^\prime}&=&
 p(k^\prime)k^\prime(k^\prime\!\!-\!1)/\bra k\ket
\end{eqnarray}
There is just one eigenvector, namely $\epsilon_k=1$ for all $k$, with eigenvalue $\lambda=\bra k^2\ket/\bra k\ket -1$.
Hence the continuous transition lines are
\begin{eqnarray}
{\rm P}\to{\rm F:}
&~~~~& 1=[\bra k^2\ket/\bra k\ket -1]\int\!\rmd J~P(J)\tanh(\beta J)
\\
{\rm P}\to{\rm SG:}
&~~~~& 1=[\bra k^2\ket/\bra k\ket -1]\int\!\rmd J~P(J)\tanh^2(\beta J)
\end{eqnarray}
The function $g(k)$ has dropped out of our equations, so the  transition lines will be identical to those of the non-deformed ensemble,
i.e. to those found when $Q(k,k^\prime)=1$.
For a Poissonian degree distribution one has $\bra k^2\ket=\bra
k\ket^2+\bra k\ket$, so $\lambda=\bra k\ket$ and we recover
the standard results for Erd\"{o}s-R\'{e}nyi graphs. Moreover, the solution $F(k)=g(k)/\bra g\ket$ of
(\ref{eq:eqnforF}) gives $Q(k,k^\prime)/F(k)F(k^\prime)=1$ for all $(k,k^\prime)$, which implies that $g(k)$ also drops out of the
order parameter equations (\ref{eq:RS1b},\ref{eq:RS2b}). It follows that for separable deformation functions $Q$
not only  the transition lines, but the complete solution
of the model, including the values of the physical observables anywhere in the phase diagram, is independent of $g(k)$ and
therefore  identical to that of
the ensemble with degree constraints only. This is true for any degree distribution, and has a simple explanation: upon substituting
$Q(k,k^\prime)=g(k)g(k^\prime)/\bra g\ket^2$ into the graph probabilities (\ref{eq:newer_connectivityB}) one obtains
\begin{eqnarray}
\hspace*{-15mm}
{\rm Prob}(\bc)&=& \frac{\rme^{N\big\{
\frac{1}{2}\bra k\ket\log [\bra k\ket/\bra g\ket^2 N]
+N^{-1}\sum_{i}k_i(\bc)\log
g(k_i(\bc))+\order(N^{-1})\Big\}}
\prod_i\delta_{k_i,k_i(\bc)}}
{\sum_{\bc^\prime}e^{N\big\{
\frac{1}{2}\bra k\ket\log [\bra k\ket/\bra g\ket^2 N]
+N^{-1}\sum_{i}k_i(\bc^\prime)\log
g(k_i(\bc^\prime))+\order(N^{-1})\Big\}}
\prod_i\delta_{k_i,k_i(\bc^\prime)}}
\nonumber
\\
\hspace*{-15mm}
&=& \frac{\rme^{\order(N^{0})}
\prod_i\delta_{k_i,k_i(\bc)}}
{\sum_{\bc^\prime}\rme^{\order(N^{0})}
\prod_i\delta_{k_i,k_i(\bc^\prime)}}
\end{eqnarray}
For $Q(k,k^\prime)=g(k)g(k^\prime)/\bra g\ket^2$
the $Q$-dependent factors in ${\rm Prob}(\bc)$ depend in leading order on $\bc$ via the degrees $k_i(\bc)$ only. Since the degrees are constrained,
these factors drop out, leaving only subdominant terms with a vanishing impact on the thermodynamics in the limit $N\to\infty$.

\subsubsection*{Type II: Additive deformation functions.}

The second class of deformation functions we will study is $Q(k,k^\prime)=[g(k)+g(k^\prime)]/2\bra g\ket$, with $g(k)\geq 0$
for all $k$ and $\bra g\ket>0$. Again the simplest choice $g(k)=1$ gives the non-deformed graph ensemble.
Now it follows from (\ref{eq:eqnforF}) that $F(k)=Ag(k)+B$, with
\begin{eqnarray}
&&
A= \frac{1}{2\bra k\ket\bra g\ket}\Big\bra\frac{k}{Ag(k)+B}\Big\ket
~~~~~~~~
B= \frac{1}{2\bra k\ket\bra g\ket}\Big\bra\frac{k g(k)}{Ag(k)+B}\Big\ket
\end{eqnarray}
We can rewrite $B$ to get a simple relation between $A$ and $B$:
\begin{eqnarray}
AB&=& \frac{1}{2\bra k\ket\bra g\ket}\Big\bra\frac{Ak g(k)}{Ag(k)+B}\Big\ket
=\frac{1}{2\bra k\ket\bra g\ket}\Big\bra k-\frac{Bk}{Ag(k)+B}\Big\ket=\frac{1}{2\bra g\ket}-AB
\end{eqnarray}
Thus $AB=1/4\bra g\ket$, i.e. $B=1/4A\bra g\ket$. Upon eliminating $B$ from our equations  and upon defining
$A=x/\bra g\ket$  with $x\geq 0$, we then find that $x$ is to be
solved from ${\cal F}(x)=1$, where
\begin{eqnarray}
{\cal F}(x)&=& \frac{2}{\bra k\ket}\Big\bra\frac{k}{1+4x^2g(k)/\bra g\ket}\Big\ket
\end{eqnarray}
We note that $d{\cal F}(x)/dx\leq 0$ for $x\geq 0$, with ${\cal
F}(0)=2$ and ${\cal F}(\infty)=0$, so there is indeed a unique and
well-defined  solution $x\geq 0$ of ${\cal F}(x)=1$.
For the trivial case $g(k)=1$ (i.e.
$Q(k,k^\prime)=1$, no ensemble deformation) we obtain ${\cal F}(x)=2/(1+4x^2)$, giving $x=\frac{1}{2}$ and the
correct simple solution $F(k)=1$ encountered earlier.

We proceed with the analysis of nontrivial choices for $g(k)$.
Let us define the short-hand $G(k)=g(k)/\bra g\ket$, so $\bra
G(k)\ket=1$,  $Q(k,k^\prime)=\frac{1}{2}[G(k)+G(k^\prime)]$, and
$F(k)=xG(k)+1/4x$.
 The matrix (\ref{eq:defineM}) to be diagonalized then takes
the following form:
\begin{eqnarray}
M_{kk^\prime}&=& \frac{8x^2}{\bra k\ket}
 \frac{[G(k)+G(k^\prime)] p(k^\prime)k^\prime(k^\prime\!\!-\!1)}{[4x^2G(k)+1][4x^2G(k^\prime)+1]}
\end{eqnarray}
Its eigenvalue equation becomes (with brackets denoting averages over the degree distribution):
\begin{eqnarray}
\hspace*{-15mm}
\lambda \epsilon_k
&=& \frac{8x^2G(k)}{\bra k\ket[4x^2G(k)\!+\!1]}\Big\bra
 \frac{k^\prime(k^\prime\!\!-\!1)\epsilon_{k^\prime}}{4x^2G(k^\prime)\!+\!1}\Big\ket
+ \frac{8x^2}{\bra k\ket[4x^2G(k)\!+\!1]}\Big\bra
 \frac{G(k^\prime) k^\prime(k^\prime\!\!-\!1)\epsilon_{k^\prime}}{4x^2G(k^\prime)\!+\!1}\Big\ket
\end{eqnarray}
We see that the components $\epsilon_k$ of any eigenvector must always be
of the form
\begin{eqnarray}
\epsilon_k&=& \frac{\cos(\phi)G(k)+\sin(\phi)}{4x^2G(k)\!+\!1}
\end{eqnarray}
where
\begin{eqnarray}
\lambda \cos(\phi)
&=& \frac{8x^2}{\bra k\ket}\Big\bra
 \frac{k(k\!-\!1)[\cos(\phi)G(k)\!+\!\sin(\phi)]}{[4x^2G(k)\!+\!1]^2}\Big\ket
\\
\lambda\sin(\phi)&=&
 \frac{8x^2}{\bra k\ket}\Big\bra
 \frac{G(k) k(k\!-\!1)[\cos(\phi)G(k)\!+\!\sin(\phi)]}{[4x^2G(k)\!+\!1]^2}\Big\ket
\end{eqnarray}
or, in matrix form:
\begin{equation}
\lambda\left(\!\!\begin{array}{c}
\cos(\phi)\\\sin(\phi)\end{array}\!\!\right)
=\frac{8x^2}{\bra k\ket}
\left(\!\begin{array}{cc}\big\bra
\frac{k(k-1)G(k)}{[4x^2G(k)+1]^2}\big\ket & \big\bra
\frac{k(k-1)}{[4x^2G(k)+1]^2}\big\ket
\\[1mm]
\big\bra \frac{k(k-1) G^2(k)}{[4x^2G(k)+1]^2}\big\ket &
 \big\bra \frac{k(k-1)
G(k)}{[4x^2G(k)+1]^2}\big\ket
\end{array}\!\right)
 \left(\!\!\begin{array}{c}
\cos(\phi)\\\sin(\phi)\end{array}\!\!\right)
\end{equation}
The two eigenvalues are now calculated easily.
We need the largest one, giving us the following expression
for $\lambda_{\rm max}(Q,p)$ for the present family of graph ensembles, with $y=4x^2/\bra g\ket$:
\begin{eqnarray}
\lambda_{\rm max}(Q,p) &=& \frac{2y}{\bra k\ket} \left\{ \Big\bra
\frac{k(k\!-\!1)g(k)}{[yg(k)\!+\!1]^2}\Big\ket+\sqrt{\Big\bra
\frac{k(k\!-\!1)}{[yg(k)\!+\!1]^2}\Big\ket~\Big\bra \frac{k(k\!-\!1)
g^2(k)}{[yg(k)\!+\!1]^2}\Big\ket} \right\}
\label{eq:eigenvalue_additive}
\end{eqnarray}
where $y$ is the solution of
\begin{eqnarray}
\Big\bra \frac{k}{yg(k)+1}\Big\ket=\frac{1}{2}\bra k\ket
\label{eq:x_additive}
\end{eqnarray}
It is a trivial matter to check that for the simple choice $g(k)=1$, where
 $y=1$,  one indeed recovers from (\ref{eq:eigenvalue_additive}) the correct eigenvalue
$\bra k^2\ket/\bra k\ket-1$ of the non-deformed ensembles.
It will also be clear from (\ref{eq:eigenvalue_additive})  that for the present non-separable family of ensemble deformation functions $Q(k,k^\prime)$, the phase diagram will
generally indeed be affected by the deformation.

\subsubsection*{Type III: Simple binary deformation functions.}

Our third class of deformation functions are those where $Q(k,k^\prime)$ takes only two values.
Here the deformation can even strictly forbid links that otherwise would have been allowed. We will focus on the  simple example
$Q(k,k^\prime)=\gamma_0+\gamma\delta_{kk^\prime}$, where  $\gamma_0=1-\gamma\sum_k p^2(k)$ and $0\leq|\gamma|\leq [\sum_k p^2(k)]^{-1}$.
The problem (\ref{eq:eqnforF}) now reduces to a quadratic equation for $F(k)$, of which the nonnegative solution is
\begin{eqnarray}
F(k)&=& \frac{1}{2}y+ \frac{1}{2}\sqrt{y^2
+4\gamma p(k)k/\bra k\ket}
\label{eq:F_binary}
\\
y &=& \frac{1\!-\!\gamma\bra p(k)\ket}{\bra k\ket} \Big\bra \frac{2k}{y + \sqrt{y^2
+4\gamma p(k)k/\bra k\ket}} \Big\ket
\label{eq:B_binary}
\end{eqnarray}
The right-hand side of (\ref{eq:B_binary}) decreases monotonically from
$[1\!-\!\gamma\bra p(k)\ket] \bra \sqrt{k/
p(k)}\ket/ \sqrt{\gamma \bra k\ket}$ at $y=0$ to zero as $y\to\infty$, so (\ref{eq:B_binary}) always has a unique non-negative solution $y$.
One also quite easily established the useful bounds
\begin{eqnarray}
\sqrt{1\!-\!\gamma\bra p(k)\ket}\Big\bra \frac{k/\bra k\ket}{\sqrt{1+4\gamma p(k)k/\bra k\ket[1\!-\!\gamma\bra p(k)\ket]}}
\leq  ~y~\leq \sqrt{1\!-\!\gamma\bra p(k)\ket}
\end{eqnarray}
(which are seen to become tight both for $\gamma\to 0$, where $y=1$, and for $\gamma\to \bra p(k)\ket^{-1}$, where $y=0$).
The matrix (\ref{eq:defineM}) will always have an eigenvalue $\lambda=0$, corresponding to the eigenspace $\epsilon_k=0$ for all $k>1$.
The eigenvectors of (\ref{eq:defineM}) with nonzero eigenvalue $\lambda$ are seen to be
\begin{eqnarray}
\epsilon_k
&=&
\frac{1}{\lambda F(k)-\gamma
  p(k)k(k\!-\!1)/\bra k\ket F(k)}
  \end{eqnarray}
  where $\lambda$ then follows upon solving
\begin{eqnarray}
1&=&
[1-\gamma\bra p(k)\ket] \Big\bra \frac{k(k\!-\!1)}{\lambda F^2(k)\bra k\ket -\gamma
  p(k)k(k\!-\!1)}\Big\ket
\end{eqnarray}
Upon inserting (\ref{eq:F_binary}) this equation takes the following explicit form, with $y$ (which is itself not a function of $\lambda$) to be solved from
(\ref{eq:B_binary}):
\begin{eqnarray}
1 &=&
[1\!-\!\gamma\bra p(k)\ket] \Big\bra \frac{4k(k\!-\!1)}{\lambda\bra k\ket[ y+ \sqrt{y^2
\!+\!4\gamma p(k)k/\bra k\ket}]^2-4\gamma p(k)k (k\!-\!1) }\Big\ket
  \label{eq:lambda_binary}
\end{eqnarray}
The right-hand side diverges to $-\infty$ for $\lambda\downarrow 0$ and
decays to zero for $\lambda\to\infty$. Furthermore, provided there exists $k>1$ with $p(k)>0$,
it has singularities for each $k$ with $p(k)>0$ at the special values $\lambda=\lambda_c(k)$, where \begin{eqnarray}
\lambda_c(k)&=& \frac{4\gamma p(k)k (k\!-\!1) }{\bra k\ket[ y+ \sqrt{y^2
\!+\!4\gamma p(k)k/\bra k\ket}]^2}\geq 0
\end{eqnarray}
 Let us define $\max_{k,p(k)>0}\lambda_c(k)=\lambda_c(k^\star)$. We know that the right-hand side of  (\ref{eq:lambda_binary})
 decreases monotonically on the interval $[\lambda_c(k^\star),\infty\ket$ from $\infty$ down to zero. Hence there is always a positive solution
 $\lambda$  of  (\ref{eq:lambda_binary}), and the largest solution $\lambda_{\rm max}(Q,p)$  lies in $[\lambda_c(k^\star),\infty\ket$.
 Furthermore,
 \begin{eqnarray}
 \lambda_{\rm max}(Q,P)&\geq & \max_k \lambda_c(k)=\max_k \frac{4\gamma p(k)k (k\!-\!1) }{\bra k\ket[ y+ \sqrt{y^2
\!+\!4\gamma p(k)k/\bra k\ket}]^2}
\label{eq:lambda_bound1}
 \end{eqnarray}
 A second simple but effective bound on solutions $\lambda>0$ is established
easily:
\begin{eqnarray}
\lambda &=&
\frac{1\!-\!\gamma\bra p(k)\ket}{\bra k\ket} \Big\bra \frac{4k(k\!-\!1)}{[ y+ \sqrt{y^2
\!+\!4\gamma p(k)k/\bra k\ket}]^2-4\gamma p(k)k (k\!-\!1)/\lambda \bra k\ket}\Big\ket
\nonumber
\\
&\geq &
\frac{1\!-\!\gamma\bra p(k)\ket}{\bra k\ket} \Big\bra \frac{4k(k\!-\!1)}{[ y+ \sqrt{y^2
\!+\!4\gamma p(k)k/\bra k\ket}]^2}\Big\ket
\geq
\frac{1\!-\!\gamma\bra p(k)\ket}{\bra k\ket} \Big\bra \frac{k(k\!-\!1)}{y^2
\!+\!4\gamma p(k)k/\bra k\ket}\Big\ket
\nonumber
\end{eqnarray}
and hence
\begin{eqnarray}
\lambda_{\rm max}(Q,P)
&\geq &
[1\!-\!\gamma\bra p(k)\ket]\Big\bra \frac{k(k\!-\!1)}{y^2\bra k\ket
\!+\!4\gamma p(k)k}\Big\ket
\label{eq:lambda_bound2}
\end{eqnarray}
For the trivial choice $\gamma=0$, where $Q(k,k^\prime)=1$, we recover the correct results for the non-deformed ensemble, viz.
$F(k)=1$ and $\lim_{\gamma\to 0}\lambda_{\rm max}(Q,p)=\bra k^2\ket/\bra k\ket-1$. Here the second bound (\ref{eq:lambda_bound2}) is
satisfied with equality.
In the opposite limit $\gamma\to \bra p(k)\ket^{-1}$, where $Q(k,k^\prime)\to \delta_{kk^\prime}/\bra p(k)\ket$, we obtain
$F(k)=\sqrt{p(k)k}/\sqrt{\bra p(k)\ket\bra k\ket}+\order(\epsilon)$ and
$y=\epsilon\bra \sqrt{k/p(k)}\ket/\sqrt{\bra k\ket/\bra p(k)\ket}+\order(\epsilon^2)$, with $\epsilon=1\!-\!\gamma\bra p(k)\ket$.
 Our equation for $\lambda$ thereby becomes
\begin{eqnarray}
\hspace*{-15mm}
1=\Big\bra \frac{k(k-1)\bra p(k)\ket}{p(k)k(\lambda\!-\!k\!+\!1)/\epsilon+\lambda\sqrt{p(k)k}\bra p(k)\ket\bra \sqrt{k/p(k)}\ket-p(k)k(\lambda
\!-\!k\!+\!1)+\order(\epsilon^2)}
\Big\ket
\end{eqnarray}
It follows that for $\gamma\to \bra p(k)\ket^{-1}$ all nonzero eigenvalues are of the form $\lambda=k^\star\!-\!1+\order(1\!-\!\gamma\bra p(k)\ket)$ with $k^\star\in\{1,2,\ldots\}$ such that $p(k^\star)>0$. The largest such eigenvalue corresponds to the largest $k^\star$ with $p(k^\star)>0$, so
$\lim_{\gamma\to \bra p(k)\ket^{-1}}\lambda_{\rm max}(Q,p)=k^\star\!-1$. Thus in the latter limit we obtain the transition lines corresponding to a regular random graph with degree $k^\star$, which is consistent with our earlier observation that for $Q(k,k^\prime)=\delta_{kk^\prime}/\bra p(k)\ket$ our graphs decompose into a collection of disconnected regular graphs, one for each degree $k$ that is allowed by $p(k)$.
Here we find that the first bound (\ref{eq:lambda_bound1}) is
satisfied with equality.

\subsection{Phase diagrams}

Our order parameter equations apply in principle to arbitrary choices of the bond distribution $P(J)$, the degree distribution $p(k)$,
and the ensemble deformation function $Q(k,k^\prime)$. Here we will limit ourselves for brevity
to the deformation functions analyzed in the previous section, and to
the binary bond distribution
$P(J)=\frac{1}{2}(1+\eta)\delta(J-J_0)+\frac{1}{2}(1-\eta)\delta(J+J_0)$, with $\eta\in[-1,1]$ and $J_0>0$.
Equations (\ref{eq:PtoF},\ref{eq:PtoSG}) can now be written as
\begin{eqnarray}
{\rm P}\to{\rm F:}
&~~~~& T_{\rm F}/J_0=2/
\log\Big[\frac{\eta\lambda_{\rm max}(Q,p)+1}{\eta\lambda_{\rm max}(Q,p)-1}\Big]
\\
{\rm P}\to{\rm SG:}
&~~~~& T_{\rm SG}/J_0=2/\log\Big[\frac{\sqrt{\lambda_{\rm max}(Q,p)}+1}{\sqrt{\lambda_{\rm max}(Q,p)}-1}\Big]
\end{eqnarray}
The P$\to$F transition is the physical one for
$\eta>\tanh(\beta J_0)$ and the P$\to$SG transition is the
physical one for $\eta<\tanh(\beta J_0)$, with a triple point
at $\eta=\tanh(\beta J_0)$.
We will consider only two types of degree distributions, both with average connectivity $\bra k\ket=c$:
\begin{eqnarray}
\hspace*{-20mm}
{\rm Poissonnian}:&~~~~p(k)=c^k\rme^{-c}/k!,&~~~~~\bra k^2\ket/c=c+1
\label{eq:poisson_degrees}
\\
\hspace*{-20mm}
{\rm power~law}: &~~~~p(k)=\Big(1\!-\!\frac{c~\zeta(3\!+\!\alpha)}{\zeta(2\!+\!\alpha)}\Big)\delta_{k0}+(1\!-\!\delta_{k0})\frac{ck^{-3-\alpha}}{\zeta(2\!+\!\alpha)},
&~~~~~\bra k^2\ket/c=\frac{\zeta(1\!+\!\alpha)}{\zeta(2\!+\!\alpha)}
\label{eq:powerlaw_degrees}
\end{eqnarray}
Here $\zeta(x)$ denotes the Riemann zeta function $\zeta(x)=\sum_{k>0}k^{-x}$ \cite{Formulas}, and we take
$\alpha\in[0,1]$ to ensure that $c=\bra k\ket$ exists (limiting ourselves to $c\leq \zeta(2\!+\!\alpha)/\zeta(3\!+\!\alpha)$, so that $p(0)\geq 0$, which means that $c$ will remain modest), but with the possibility to take the scale-free limit $\alpha\to 0$.
In practice, however, in calculating averages over $p(k)$ numerically one has to truncate the values of $k$; here we used $k\leq k_{\rm max}= 10^8$.
For Poissonnian $p(k)$ this has no noticeable implications, but for power law $p(k)$ the slow divergence of $\sum_k k^{-1}\approx \log k_{\rm max}$ manifests itself in transition temperatures for $\alpha\to 0$ that should have been infinite but are finite.
On the other hand, in any finite real system or simulation one will have $k<N$, so one expects to see also there
 the same effects of bounded degrees (e.g. finite transition temperatures). The `ideal' situation of unbounded degrees and truly scale-free graphs is never realized in practice. The power-law distribution (\ref{eq:powerlaw_degrees}) has the property $p(k)|_c=cp(k)_{c=1}$ for $k>0$.
  Hence for any function $\psi(k)$ with $\psi(0)=0$ one will have $\bra \psi(k)\ket=c\bra \psi(k)\ket_{c=1}$. As a consequence one finds immediately upon checking the various formulae of the previous section that the bifurcation lines for type I and type II ensemble deformations
 are completely independent of the connectivity $c$ for power-law distributed degrees. For type III deformations this is not the case.
Note, finally, that
there is no point in choosing regular graphs $p(k)=\delta_{kc}$,
since there the function $Q(k,k^\prime)$ is always equal to one due to the normalization requirement
$\sum_{kk^\prime}p(k)p(k^\prime)Q(k,k^\prime)=1$.

\begin{figure}[t]
\vspace*{7mm} \hspace*{-17mm} \setlength{\unitlength}{0.62mm}
\begin{picture}(200,80)
 \put(64,9){\here{$c$}} \put(154,9){\here{$c$}}  \put(244,9){\here{$c$}}
 \put(20,55){\here{\large $\frac{T}{J_0}$}} \put(110,55){\here{\large $\frac{T}{J_0}$}} \put(200,55){\here{\large $\frac{T}{J_0}$}}
\put(25,15){\includegraphics[width=105\unitlength]
{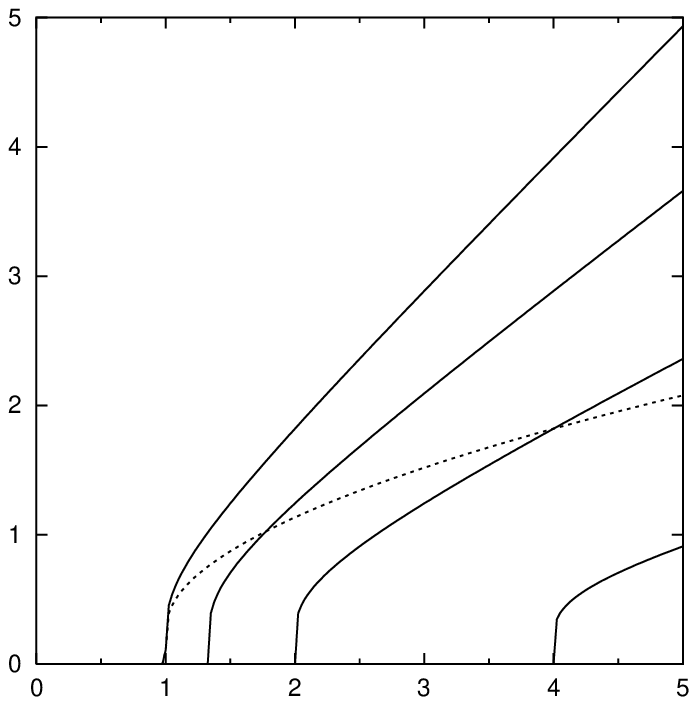}}
\put(115,15){\includegraphics[width=105\unitlength]
{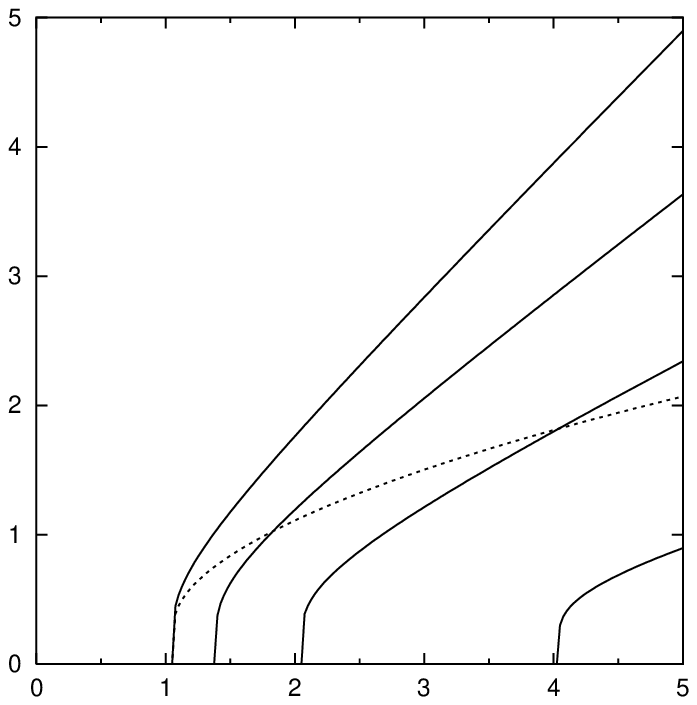}}
\put(205,15){\includegraphics[width=105\unitlength]
{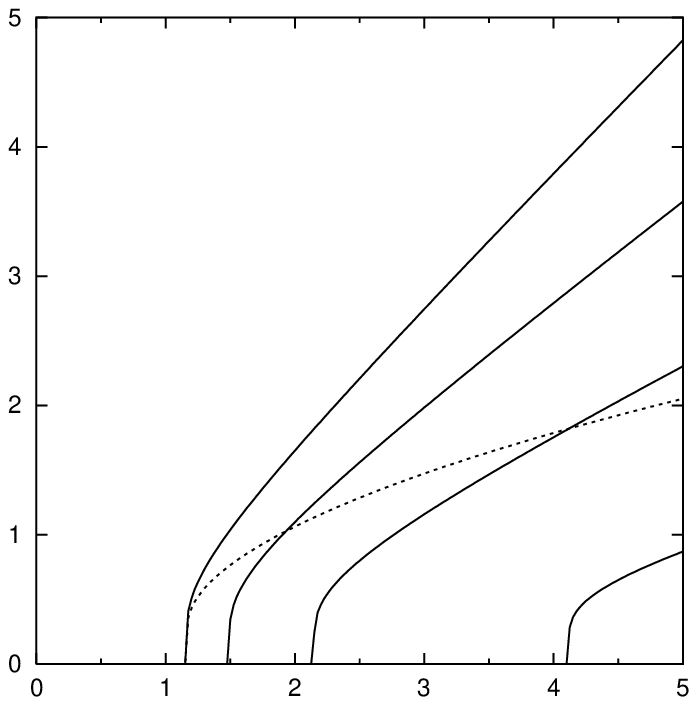}}
\put(35,77){\small\sl $g(k)=1$}  \put(125,77){\small\sl $g(k)=k$}  \put(215,77){\small\sl $g(k)=k^2$}
\end{picture}
\vspace*{-5mm} \caption{Continuous bifurcation lines for P$\to$SG (dotted) and P$\to$F (solid, with $\eta\in\{0.25,0.5,0.75,1\}$ from bottom to top), for type II deformed ensembles (with $Q(k,k^\prime)=[g(k)+g(k^\prime)]/2\bra g(k)\ket$),  $P(J)=\frac{1}{2}(1+\eta)\delta(J-J_0)+\frac{1}{2}(1-\eta)\delta(J+J_0)$, and Poissonnian degree distributions $p(k)=c^k\rme^{-c}/k!$.
Left to right: $g(k)\in\{1,k,k^2\}$. The left picture represents the non-deformed ensemble, to serve as a reference. The effect of a deformation with $Q(k,k^\prime)=[k^m+(k^\prime)^m]/2\bra k^m\ket$ in graphs with Poissonnian $p(k)$ is seen to be a slight reduction of all critical temperatures with increasing $m$.
 }\label{fig:defIIpoisson}
\end{figure}

\begin{figure}[t]
\vspace*{7mm} \hspace*{-17mm} \setlength{\unitlength}{0.62mm}
\begin{picture}(200,80)
 \put(64,9){\here{$\alpha$}} \put(154,9){\here{$\alpha$}}  \put(244,9){\here{$\alpha$}}
 \put(20,55){\here{\large $\frac{T}{J_0}$}} \put(110,55){\here{\large $\frac{T}{J_0}$}} \put(200,55){\here{\large $\frac{T}{J_0}$}}
\put(25,15){\includegraphics[width=105\unitlength]
{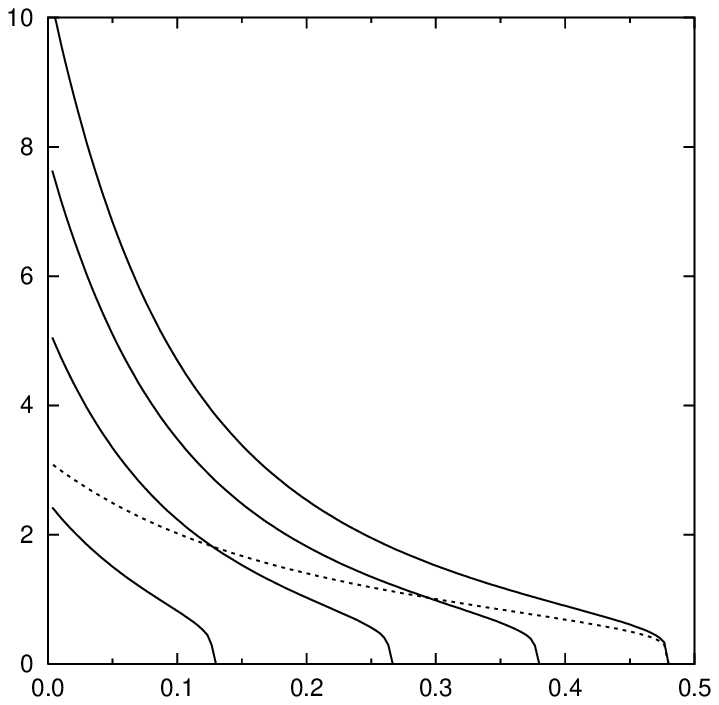}}
\put(115,15){\includegraphics[width=105\unitlength]
{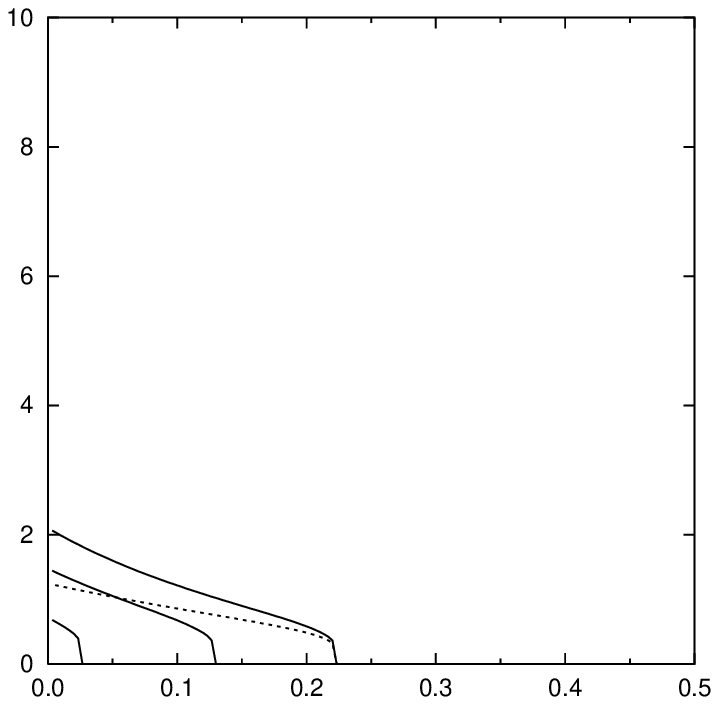}}
\put(205,15){\includegraphics[width=105\unitlength]
{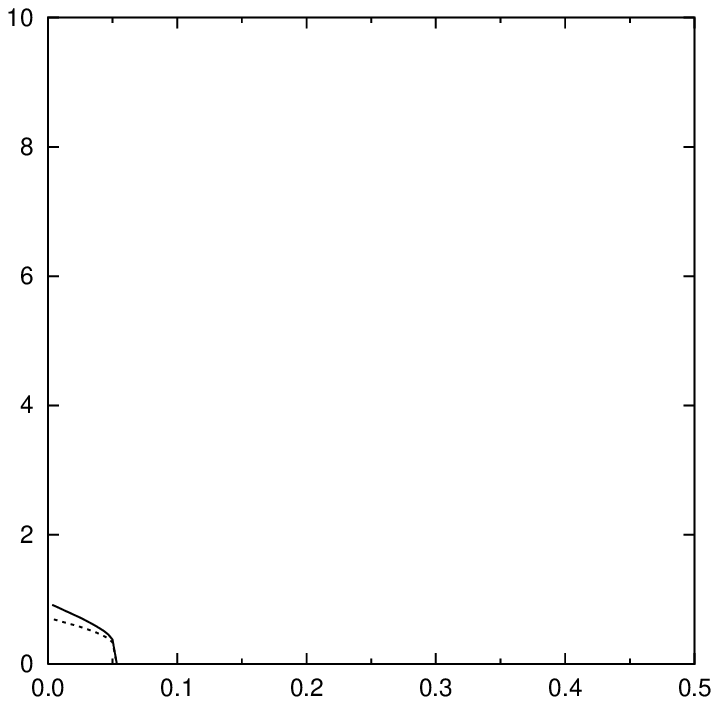}}
\put(68,77){\small\sl $g(k)=1$}  \put(158,77){\small\sl $g(k)=k$}  \put(248,77){\small\sl $g(k)=k^2$}
\end{picture}
\vspace*{-5mm} \caption{Continuous bifurcation lines for P$\to$SG (dotted) and P$\to$F (solid, with $\eta\in\{0.25,0.5,0.75,1\}$ from bottom to top), for type II deformed ensembles (with $Q(k,k^\prime)=[g(k)+g(k^\prime)]/2\bra g(k)\ket$),  $P(J)=\frac{1}{2}(1+\eta)\delta(J-J_0)+\frac{1}{2}(1-\eta)\delta(J+J_0)$, and power law degree distributions $p(k)\sim k^{-3-\alpha}$.
Left to right: $g(k)\in\{1,k,k^2\}$. The left picture represents the non-deformed ensemble, to serve as a reference.
The effect of a deformation with $Q(k,k^\prime)=[k^m+(k^\prime)^m]/2\bra k^m\ket$ in graphs with power law $p(k)$ is now seen to be a dramatic reduction of all critical temperatures with increasing $m$.
 }\label{fig:defIIpowerlaw}
\end{figure}

We will compare phase diagrams for the previously analyzed families of deformation functions $Q(k,k^\prime)$, viz. the
separable ones,
the additive ones, and the binary ones. In the separable case (type I), where one always has the simple eigenvalue $\lambda_{\rm max}(Q,p)=\bra k^2\ket/\bra k\ket-1$, we have fully explicit expressions for the transition lines that are identical to those describing
non-deformed ensembles with degree constraints only:
\begin{eqnarray}
\hspace*{-15mm}
{\rm Poissonnian~}p(k):
&~~~&T_{\rm F}/J_0=2/
\log\Big[\frac{\eta c+1}{\eta c-1}\Big],~~~~~~
T_{\rm SG}/J_0=2/\log\Big[\frac{\sqrt{c}+1}{\sqrt{c}-1}\Big]
\label{eq:nondef_poisson_lines}
\\
\hspace*{-15mm}
{\rm power~law~}p(k):
&~~~~& T_{\rm F}/J_0=2/
\log\left[\frac{\eta\zeta(1\!+\!\alpha)+(1\!-\!\eta)\zeta(2\!+\!\alpha)}
{\eta\zeta(1\!+\!\alpha)-(1\!+\!\eta)\zeta(2\!+\!\alpha)}\right],
\label{eq:nondef_power_linesa}
\\
\hspace*{-15mm}
&~~~~& T_{\rm SG}/J_0=2/\log\Big[
\frac{\sqrt{\zeta(1\!+\!\alpha)-\zeta(2\!+\!\alpha)}+\sqrt{\zeta(2\!+\!\alpha)}}
{\sqrt{\zeta(1\!+\!\alpha)-\zeta(2\!+\!\alpha)}-\sqrt{\zeta(2\!+\!\alpha)}}\Big]
\label{eq:nondef_power_linesb}
\end{eqnarray}
(in non-deformed graphs of the type considered here, the transition temperatures for power-law
distributed degree distributions are independent of the average connectivity).
Clearly, in non-deformed Poissonnian graphs we can only have an SG phase if $c>1$ and an $F$ phase if $c>1/\eta$, whereas
in non-deformed power law graphs   we can only have an SG phase if $\zeta(1\!+\!\alpha)/\zeta(2\!+\!\alpha)>2$ (giving $\alpha<\alpha_c\approx 0.479$)  and an $F$ phase if $\zeta(1\!+\!\alpha)/\zeta(2\!+\!\alpha)>1+1/\eta$. We will not show these lines describing the non-deformed ensembles
 in a separate figure, but will
include them as a benchmark when showing data for the type II and type III deformations, since in type II models the non-deformed ensemble is recovered
for the special choice $g(k)=1$ whereas in the type III models it corresponds to $\gamma=0$.

\begin{figure}[t]
\vspace*{7mm} \hspace*{-17mm} \setlength{\unitlength}{0.62mm}
\begin{picture}(200,80)
 \put(64,9){\here{$c$}} \put(154,9){\here{$c$}}  \put(244,9){\here{$c$}}
 \put(20,55){\here{\large $\frac{T}{J_0}$}} \put(110,55){\here{\large $\frac{T}{J_0}$}} \put(200,55){\here{\large $\frac{T}{J_0}$}}
\put(25,15){\includegraphics[width=105\unitlength]
{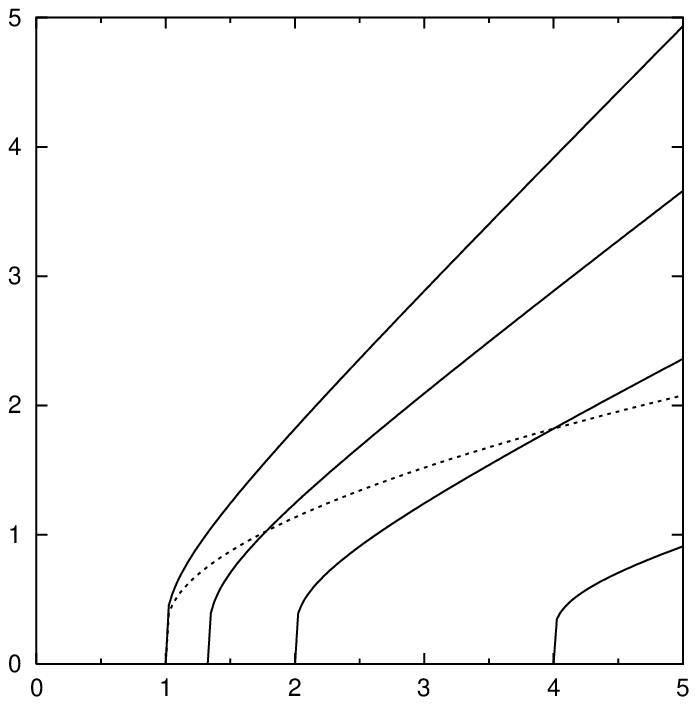}}
\put(115,15){\includegraphics[width=105\unitlength]
{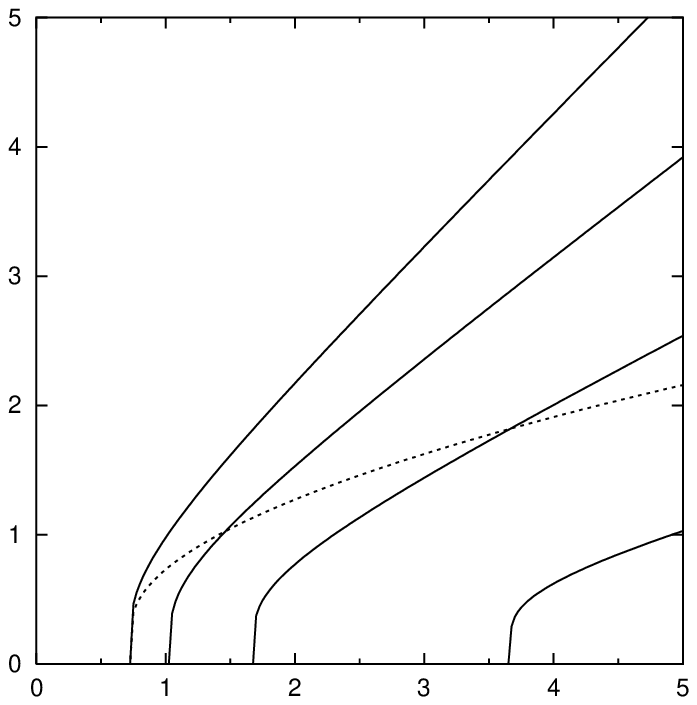}}
\put(205,15){\includegraphics[width=105\unitlength]
{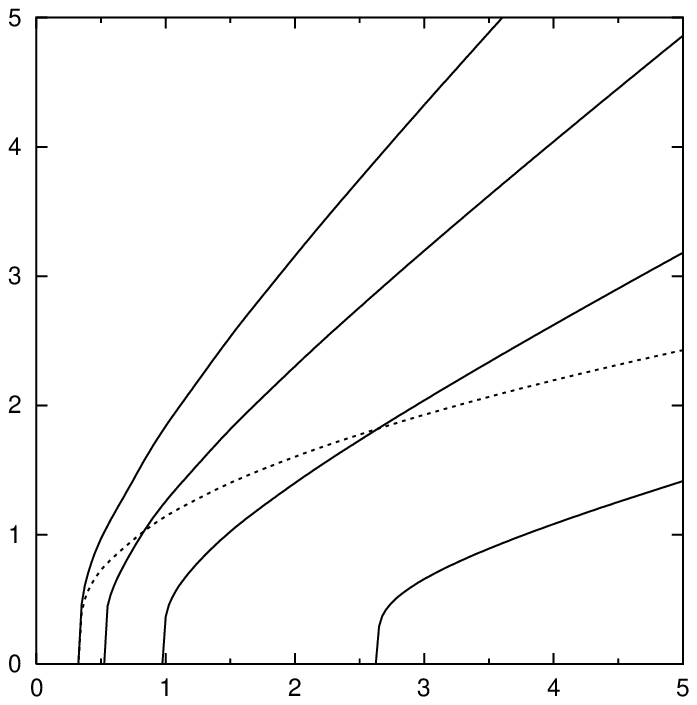}}
\put(35,77){\small\sl $\gamma=0$}  \put(124,77){\small\sl $\gamma\!=\!\frac{2}{5}\bra p(k)\ket^{-1}$}
\put(212,77){\small\sl $\gamma\!=\!\frac{4}{5}\bra p(k)\ket^{-1}$}
\end{picture}
\vspace*{-5mm} \caption{Continuous bifurcation lines for P$\to$SG (dotted) and P$\to$F (solid, with $\eta\in\{0.25,0.5,0.75,1\}$ from bottom to top), for type III deformed ensembles (with $Q(k,k^\prime)=\gamma_0+\gamma \delta_{kk^\prime}$),  $P(J)=\frac{1}{2}(1+\eta)\delta(J-J_0)+\frac{1}{2}(1-\eta)\delta(J+J_0)$, and Poissonnian degree distributions $p(k)=c^k\rme^{-c}/k!$.
Left to right: $\gamma\bra p(k)\ket\in\{0,0.4,0.8\}$. The left picture represents the non-deformed ensemble, to serve as a reference. The effect of a deformation with $Q(k,k^\prime)=\gamma_0+\gamma \delta_{kk^\prime}$ in graphs with Poissonnian $p(k)$ is seen to be a significant increase of all critical temperatures with increasing $\gamma$.
 }\label{fig:defIIIpoisson}
\end{figure}

\begin{figure}[t]
\vspace*{7mm} \hspace*{-17mm} \setlength{\unitlength}{0.62mm}
\begin{picture}(200,80)
 \put(64,9){\here{$c$}} \put(154,9){\here{$c$}}  \put(244,9){\here{$c$}}
 \put(20,55){\here{\large $\frac{T}{J_0}$}} \put(110,55){\here{\large $\frac{T}{J_0}$}} \put(200,55){\here{\large $\frac{T}{J_0}$}}
\put(25,15){\includegraphics[width=105\unitlength]
{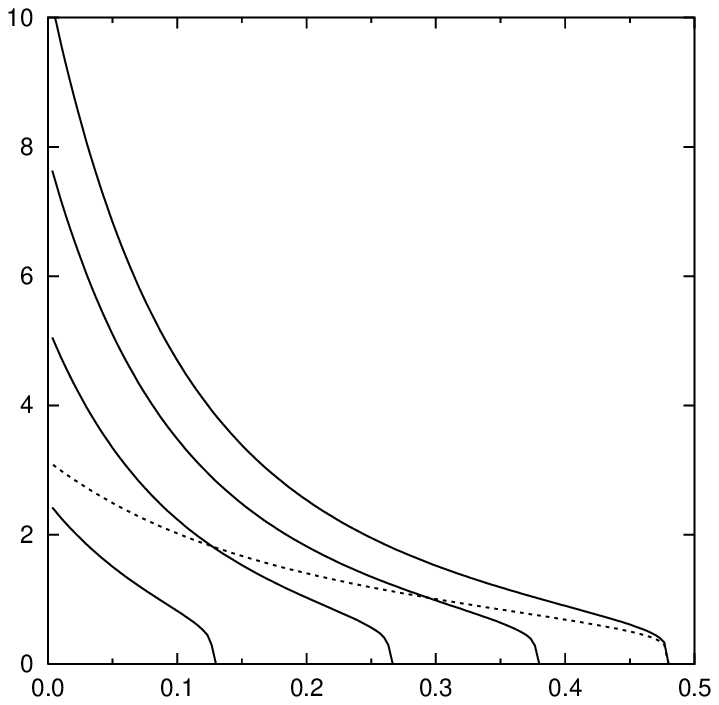}}
\put(115,15){\includegraphics[width=105\unitlength]
{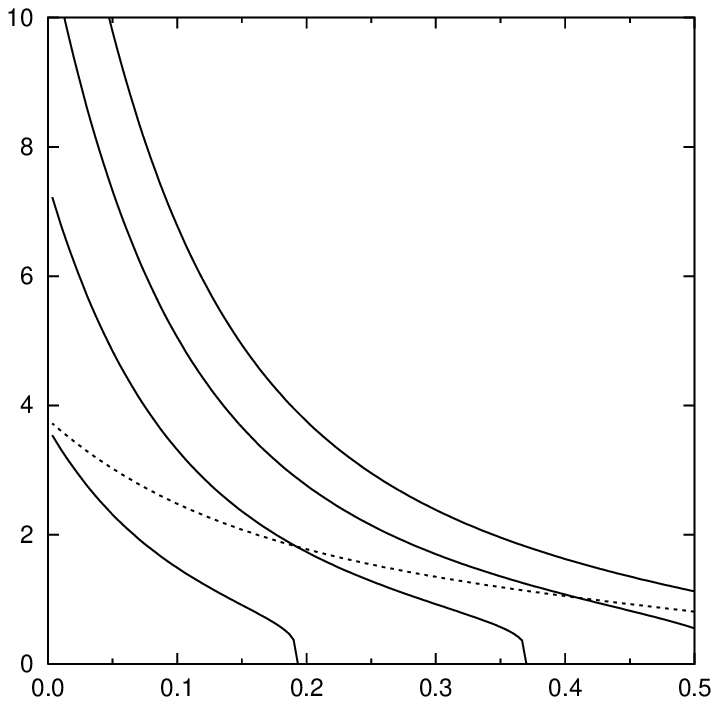}}
\put(205,15){\includegraphics[width=105\unitlength]
{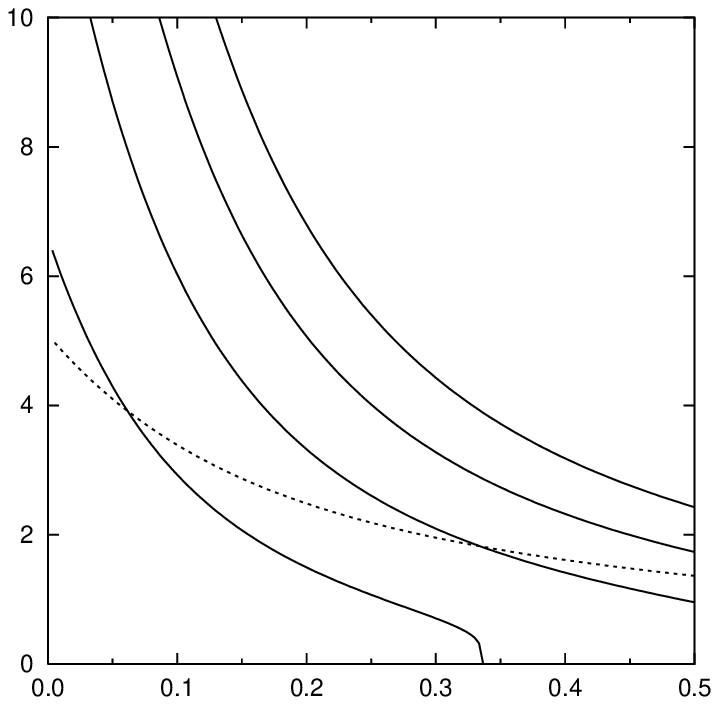}}
\put(80,77){\small\sl $\gamma=0$}  \put(151,77){\small\sl $\gamma\!=\!\frac{2}{5}\bra p(k)\ket^{-1}$}  \put(241,77){\small\sl $\gamma\!=\!\frac{4}{5}\bra p(k)\ket^{-1}$}
\end{picture}
\vspace*{-5mm} \caption{Continuous bifurcation lines for P$\to$SG (dotted) and P$\to$F (solid, with $\eta\in\{0.25,0.5,0.75,1\}$ from bottom to top), for type III deformed ensembles (with $Q(k,k^\prime)=\gamma_0+\gamma\delta_{kk^\prime}$),  $P(J)=\frac{1}{2}(1+\eta)\delta(J-J_0)+\frac{1}{2}(1-\eta)\delta(J+J_0)$, and power law degree distributions $p(k)\sim k^{-3-\alpha}$ with $c=1$.
Left to right: $\gamma\bra p(k)\ket\in\{0,0.4,0.8\}$. The left picture represents the non-deformed ensemble, to serve as a reference.
The effect of a deformation with $Q(k,k^\prime)=\gamma_0+\gamma\delta_{kk^\prime}$ in graphs with power law $p(k)$ is now seen to be a dramatic
 increase of all critical temperatures with increasing $\gamma$.
 }\label{fig:defIIIpowerlaw}
\end{figure}

For the  additive functions $Q(k,k^\prime)$ (type II) the eigenvalue $\lambda_{\rm max}(Q,p)$ depends in a nontrivial way on
$Q(k,k^\prime)$ and $p(k)$, and must be solved from
(\ref{eq:eigenvalue_additive},\ref{eq:x_additive}) numerically.
Here we will choose either $g(k)=k$ or $g(k)=k^2$ in the function $Q(k,k^\prime)$. For Poissonnian distributed
degrees this gives $G(k)=k/c$ and $G(k)=k^2/c(c+1)$ respectively.
For the power law distributed degrees one finds $G(k)=k/c$ and $G(k)=k^2\zeta(2\!+\!\alpha)/c\zeta(1\!+\!\alpha)$, respectively.
Upon solving (\ref{eq:eigenvalue_additive},\ref{eq:x_additive}) numerically for Poissonnian graphs, we obtain the bifurcation lines as
shown in Figure \ref{fig:defIIpoisson}. We also show the lines for the non-deformed case $g(k)=1$, as a benchmark.
The deformation causes only minor changes to the phase diagram, mainly a slight reduction
of all transition temperatures for small values of the connectivity $c$. When applied to graphs with power law degrees, in contrast, the impact
of the deformation is much more drastic, as shown in Figure \ref{fig:defIIpowerlaw}.
This can be understood mathematically on the basis of equations (\ref{eq:eigenvalue_additive},\ref{eq:x_additive}). If we consider the case $\alpha\to 0$ we only need to inspect what happens to the divergent sums over $k$: one finds for both $g(k)=k$ and $g(k)=k^2$ that
 $\lambda_{\rm max}(Q,p)\sim \sqrt{\log k_{\rm max}}$ as $\alpha\downarrow 0$ (rather than
$\lambda_{\rm max}(Q,p)\sim \log k_{\rm max}$, as was the case for the non-deformed ensemble).

Finally, we have solved numerically  equations (\ref{eq:B_binary},\ref{eq:lambda_binary}) for the case of the binary functions $Q(k,k^\prime)$ (type III), for $\gamma\bra p(k)\ket\in\{0,0.4,0.8\}$ (with the first value $\gamma=0$, the non-deformed case, serving as a benchmark) and $c=1$. Positive values of $\gamma$ imply increased connections between links with identical degree, which favours especially the formation regular graphs with large values of $k$. Here we
always observe a significant increase of all critical temperatures, both for Poissonnian and for power-law distributed graphs, see Figure \ref{fig:defIIIpowerlaw}. The effect becomes stronger as $c$ increases.
Choosing negative values of $\gamma$, i.e. discouraging the formation of links between nodes with identical degree, is found to decrease all transition temperatures.
Note that without the degree cut-off $k_{\rm max}=10^8$, one would have diverging critical temperatures at all $c>0$ and all $\alpha\geq 0$ in the limit $\gamma\bra p(k)\ket\to 1$.

\section{Discussion}

The rationale behind studying interacting particle  models on complex {\em random} graphs is that the latter can be used as solvable proxies for  models on {\em specific} graphs of a topology for which either no exact solution is available (e.g. spin models on cubic lattices), or on which we
lack precise information (e.g. proteomic networks). We then have to choose an appropriate ensemble of random graphs, which is sufficiently simple to allow for analytical progress, while incorporating as much as possible of the topology of the specific system one aims to understand.
 Specifying just the degree distribution $p(k)$
of a complex connectivity graph for an interacting spin system will clearly not yet permit reliable predictions on the system's phase diagram.
For instance, the critical temperature of the D-dimensional Ising model on a cubic lattice, where $p(k)=\delta_{k,2D}$, is different from that of a regular random graph with $p(k)=\delta_{k,2D}$\footnote{For a ferromagnetic  Ising model on a square lattice in $D=2$ one has Onsager's famous result $T_c/J_0=4/\log[(\sqrt{2}+1)/(\sqrt{2}-1)]\approx 2.26919$, whereas $T_c/J_0=2/\log 2\approx 2.88539$ in the degree-4 regular random graph. Yet both models have the same degree distribution $p(k)=\delta_{k,4}$.}.
The question is then which further topological information on a graph beyond $p(k)$ could be added
to  reduce the entropy of the underlying graph ensemble and make more specific and more accurate predictions of phase transitions, while at the same time maintaining the vital property that the resulting spin models can be solved analytically.  In this paper we have established that
the proposed deformation of random graph ensembles can be a useful step in this direction: it generally allows us to differentiate between
models with the same $p(k)$ (which can be chosen freely) but different microscopic realizations of these degree statistics,
the resulting models are still solvable, and its impact on the transition lines can be non-negligible\footnote{The only graphs where ensemble deformation is not possible are the regular graphs, with $p(k)=\delta_{kc}$, where the constraint $\sum_{kk^\prime}p(k)Q(k,k^\prime)p(k^\prime)=1$ leaves only the trivial choice $Q(k,k^\prime)=1$.}.
In practice, when seeking to model a complicated real system with some specific given interaction graph $\bc^\star$ (and hence a known
 set of degree $\{k^\star_1,\ldots,k_N^\star\}$ and a known degree distribution $p(k)$)
 by a solvable system on a random graph, we could now incorporate at least some of the extra topological information
by using our ensemble (\ref{eq:newer_connectivity}) with constrained degrees $k_i=k_i^\star$ for all $i$, and with
a function $Q(k,k^\prime)$ that is taylored to the graph $\bc^\star$. This can be done by maximizing the log-likelihood of $\bc^\star$ for the ensemble (\ref{eq:newer_connectivity}), i.e. by minimizing
over $Q$ (subject to $\sum_{kk^\prime}p(k)p(k^\prime)Q(k,k^\prime)=1$) the quantity
\begin{eqnarray}
\hspace*{-15mm}
\Omega[Q]&=& \frac{1}{N}\log {\mathcal Z}_{N} -\frac{1}{N}\sum_{i<j}\log \left[\frac{\bra k\ket
}{N}Q(k^\star_i,k^\star_j)\delta_{c^\star_{ij},1}+\Big(1\!-\!\frac{\bra k\ket
}{N}Q(k^\star_i,k^\star_j)\Big)\delta_{c^\star_{ij},0}\right]
\nonumber
\\
\hspace*{-15mm}
&=& z +\frac{1}{2}\bra k\ket -\frac{1}{2}\bra k\ket \log [\bra k\ket
/N]
-\frac{1}{N}\sum_{i<j}c_{ij}^\star
\log Q(k^\star_i,k^\star_j)
+\order(N^{-1})
\nonumber
\\
\hspace*{-15mm}
&=& {\rm const}~
+\sum_{k} p(k)k\log F(k|Q)
-\frac{1}{N}\sum_{i<j}c_{ij}^\star
\log Q(k^\star_i,k^\star_j)
+\order(N^{-1})
\end{eqnarray}
where $F(k|Q)$ is the solution of
\begin{eqnarray}
F(k)= \bra k\ket^{-1}\sum_{k^\prime}p(k^\prime)k^\prime Q(k,k^\prime) F^{-1}(k^\prime)
\end{eqnarray}
This will be the subject of a subsequent study. In addition one would like to study certain technical aspects
of the present model in more detail, such as the precise physical meaning of the function $F(k)$, and the
impact of possible replica symmetry breaking (RSB). In the present type of model RSB does not change the locations of the
P$\to$F or P$\to$SG transition lines, but will alter the nature of the solution in the ordered phases and the location of the F$\to$SG transition line.

\section*{Acknowledgements}

It is our great pleasure to thank Ginestra Bianconi and Isaac Perez-Castillo for valuable discussions.
One of the the authors (CJPV) acknowledges financial
support from project FIS2006-13321-C02-01 and grant PR2006-0458.

\appendix

\section{Joint distributions of degree and clustering coefficients}
\label{app:clustering}

To characterized a graph's local topology
we can define for each vertex $i$ the degree
$k_i(\bc)=\sum_{j}c_{ij}$ (the number of links to this vertex)
and the number of length-three loops going through this vertex, as measured by
$r_i(\bc)=\sum_{jk}c_{ij}c_{jk}c_{ki}$. The clustering coefficient $C_i$ is then given by $C_i=r_i/k_i(k_i-1)$.
We write their joint
distribution as $P(k,r|\bc)=N^{-1}\sum_i \delta_{k,k_i(\bc)}\delta_{r,r_i(\bc)}$, and the asymptotic expectation value of this distribution
over the ensemble
(\ref{eq:newer_connectivity}) as
  \begin{eqnarray}
P(k,r)&=& \lim_{N\to\infty}\frac{1}{N{\mathcal Z}_N}\sum_i\sum_{\bc}
\delta_{k,k_i}\delta_{r,r_i}
 = \int\!\frac{\rmd\psi}{2\pi}\rme^{\rmi\psi r}
\lim_{N\to\infty}\frac{1}{N}\sum_i \delta_{k,k_i}\hat{P}_i(\psi)
\end{eqnarray}
with \begin{eqnarray}
\hat{P}_i(\psi)&=&  \frac{1}{{\mathcal Z}_N}\sum_{\bc}
\rme^{-\rmi\psi\sum_{j\ell}c_{ij}c_{j\ell}c_{\ell i}}
\end{eqnarray}
It will turn out that here  we have to expand to higher orders in $N$ than in previous calculations.
 We now
re-name all links to/from site $i$ as
$s_j=c_{ij}\in\{0,1\}$, while writing all those that do not
involve site $i$ as $\tau_{j\ell}\in\{0,1\}$, where
$j,\ell\in\{1,\ldots,i-1,i+1,\ldots,N\}$. This gives
\begin{eqnarray}
\hspace*{-20mm}
 \hat{P}_i(\psi)&=& \frac{1}{\mathcal
Z_N}\sum_{\bes \btau} \rme^{-\rmi\psi\sum_{j\ell\neq
i}s_{j}\tau_{j\ell}s_{\ell }} \prod_{j\neq
i}\left[\frac{\bra k\ket}{N}Q(k_i,k_j)\delta_{s_{j},1}+\Big(1\!-\!\frac{\bra k\ket}{N}Q(k_i,k_j)\Big)\delta_{s_{j},0}\right]\delta_{k_i,\sum_{j\neq
i} s_{j}}\nonumber
\\
\hspace*{-20mm}
 &&\hspace*{20mm}
 \times\!\!
\prod_{\ell<j|\ell,j\neq
i}\left[\frac{\bra k\ket}{N}Q(k_\ell,k_j)\delta_{\tau_{\ell
j},1}+\Big(1\!-\!\frac{\bra k\ket}{N}Q(k_\ell,k_j)\Big)\delta_{\tau_{\ell
j},0}\right]\prod_{\ell\neq i}\delta_{k_\ell,\sum_{j\neq i,\ell}
\tau_{\ell j}} \nonumber
\\
\hspace*{-20mm} &=&
 \frac{1}{\mathcal
Z_N}\sum_{\bes}\int_{-\pi}^{\pi}\!\frac{\rmd\phi}{2\pi}
\rme^{\rmi\phi k_i} \prod_{j\neq
i}\left[\frac{\bra k\ket}{N}Q(k_i,k_j)\delta_{s_{j},1}+\Big(1\!-\!\frac{\bra k\ket}{N}Q(k_i,k_j)\Big)\delta_{s_{j},0}\right]\rme^{-\rmi\phi
\sum_{j\neq i} s_{j} } \nonumber
\\
\hspace*{-25mm}
 &&\times\int_{-\pi}^{\pi}\!\prod_{\ell\neq
 i}\Big[\frac{\rmd\phi_\ell}{2\pi}\rme^{\rmi\phi_\ell k_\ell}\Big]
\rme^{\frac{\bra k\ket}{2N}\sum_{\ell j (\neq i)}Q(k_\ell,k_j)[ \rme^{-\rmi(2\psi
s_{\ell }s_j+\phi_\ell+\phi_j)}-1 ]+\order(N^{-1})} \nonumber
\\
\hspace*{-25mm}&&\times \rme^{-\frac{\bra k\ket^2}{4N^2}\sum_{\ell j (\neq
i)}Q^2(k_\ell,k_j)[ \rme^{-\rmi(2\psi s_{\ell
}s_j+\phi_\ell+\phi_j)}-1]^2  -\frac{\bra k\ket}{2N}\sum_{j\neq
i}Q(k_j,k_j) [\rme^{-2\rmi(\psi s^2_j+\phi_j)}-1] }
\end{eqnarray}
At this point we are led to the introduction of the observables
\begin{equation}
W_{\!s k}(\phi)=\frac{1}{N\!-\!1}\sum_{j\neq i}\delta_{s
s_j}\delta_{k k_j}\delta(\phi\!-\!\phi_j)
\end{equation}
Clearly $\sum_{s\in\{0,1\}}\sum_{k\geq 0} \int\!d\phi~W_{\!sk}(\phi)=1$.
We also introduce the short-hand $p_k=N^{-1}\sum_i\delta_{k,k_i}$ (viz. the empirical degree frequencies,
which will only be identical to $p(k)$ for $N\to\infty$).
Upon introducing the $W_{sk}(\phi)$ in the usual manner via
suitable $\delta$-functions we can then write
\begin{eqnarray}
\hspace*{-20mm}
 \hat{P}_i(\psi)&=&
 \frac{1}{\mathcal
Z_c}\int_{-\pi}^{\pi}\!\frac{\rmd\phi}{2\pi} \rme^{\rmi\phi k_i}\int\!\{\rmd W
\rmd\hat{W}\}~\rme^{\rmi(N-1)\sum_{s^\prime k^\prime}\int\!\rmd\phi^\prime
\hat{W}_{\!s^\prime k^\prime}(\phi^\prime) W_{\!s^\prime
k^\prime}(\phi^\prime)+\order(N^{-1})} \nonumber
\\
\hspace*{-25mm} && \times \rme^{\frac{1}{2}(N-2)\bra k\ket \sum_{s^\prime k^\prime
s^\pprime k^\pprime}\int\!\rmd\phi^\prime \rmd\phi^\pprime W_{\!s^\prime
k^\prime}(\phi^\prime) W_{\!s^\pprime k^\pprime}(\phi^\pprime)
Q(k^\prime,k^\pprime)[ \rme^{-\rmi(2\psi s^\prime
s^\pprime+\phi^\prime+\phi^\pprime)}-1 ]} \nonumber
\\
\hspace*{-25mm}&&\times \rme^{-\frac{1}{4}\bra k\ket^2\sum_{s^\prime k^\prime
s^\pprime k^\pprime }\int\!\rmd\phi^\prime d\phi^\pprime
W_{\!s^\prime k^\prime}(\phi^\prime) W_{\!s^\pprime
k^\pprime}(\phi^\pprime) Q^2(k^\prime,k^\pprime)[ \rme^{-\rmi(2\psi
s^\prime s^\pprime+\phi^\prime+\phi^\pprime)}-1]^2} \nonumber
\\
\hspace*{-25mm} &&\times \rme^{-\frac{1}{2}\bra k\ket\sum_{s^\prime
k^\prime}\int\!\rmd\phi^\prime W_{\!s^\prime k^\prime}(\phi^\prime)
Q(k^\prime,k^\prime) [\rme^{-2\rmi(\psi s^{\prime}+\phi^\prime)}-1] }
\nonumber
\\
\hspace*{-25mm} &&
 \times \prod_{j\neq i}\Big\{\sum_{s}\int_{-\pi}^{\pi}\!\frac{\rmd\phi^\prime}{2\pi}
\Big[\frac{\bra k\ket}{N}Q(k_i,k_j)\delta_{s,
1}+\Big(1\!-\!\frac{\bra k\ket}{N}Q(k_i,k_j)\Big)\delta_{s,0}\Big]\rme^{-\rmi\phi
s+\rmi[\phi^\prime
 k_j-\hat{W}_{s k_j}(\phi^\prime)]}\Big\}
 \nonumber\\
 \hspace*{-20mm}
&=&
\frac{1}{\mathcal
Z_c}\int_{-\pi}^{\pi}\!\frac{\rmd\phi}{2\pi} \rme^{\rmi\phi k_i}\int\!\{\rmd W
\rmd\hat{W}\}~\rme^{\rmi(N-1)\sum_{s^\prime k^\prime}\int\!\rmd\phi^\prime
\hat{W}_{\!s^\prime k^\prime}(\phi^\prime) W_{\!s^\prime
k^\prime}(\phi^\prime)+\order(N^{-1})} \nonumber
\\
\hspace*{-25mm} && \times \rme^{\frac{1}{2}(N-2)\bra k\ket \sum_{s^\prime k^\prime
s^\pprime k^\pprime}\int\!\rmd\phi^\prime \rmd\phi^\pprime W_{\!s^\prime
k^\prime}(\phi^\prime) W_{\!s^\pprime k^\pprime}(\phi^\pprime)
Q(k^\prime,k^\pprime)[ \rme^{-\rmi(2\psi s^\prime
s^\pprime+\phi^\prime+\phi^\pprime)}-1 ]} \nonumber
\\
\hspace*{-25mm}&&\times \rme^{-\frac{1}{4}\bra k\ket^2\sum_{s^\prime k^\prime
s^\pprime k^\pprime }\int\!\rmd\phi^\prime d\phi^\pprime
W_{\!s^\prime k^\prime}(\phi^\prime) W_{\!s^\pprime
k^\pprime}(\phi^\pprime) Q^2(k^\prime,k^\pprime)[ \rme^{-\rmi(2\psi
s^\prime s^\pprime+\phi^\prime+\phi^\pprime)}-1]^2} \nonumber
\\
\hspace*{-25mm} &&\times \rme^{-\frac{1}{2}\bra k\ket\sum_{s^\prime
k^\prime}\int\!\rmd\phi^\prime W_{\!s^\prime k^\prime}(\phi^\prime)
Q(k^\prime,k^\prime) [\rme^{-2\rmi(\psi s^{\prime}+\phi^\prime)}-1] }
\nonumber
\\
\hspace*{-25mm} &&\hspace*{-10mm}
 \times \prod_{j\neq i}\left\{
\Big[ \rme^{-\frac{\bra k\ket}{N}Q(k_i,k_j)}\int_{-\pi}^{\pi}\!\frac{\rmd\phi^\prime}{2\pi}\rme^{\rmi[\phi^\prime
 k_j-\hat{W}_{0 k_j}(\phi^\prime)]}
\Big]\Big[ 1\!+\! \frac{\bra k\ket}{N}Q(k_i,k_j)\frac{
 \int_{-\pi}^{\pi}\!\rmd\phi^\prime \rme^{\rmi[\phi^\prime
 k_j-\hat{W}_{1 k_j}(\phi^\prime)-\phi]}}
 { \int_{-\pi}^{\pi}\!\rmd\phi^\prime \rme^{\rmi[\phi^\prime
 k_j-\hat{W}_{0 k_j}(\phi^\prime)]}}
\Big]\right\}
 \nonumber\\
 \hspace*{-25mm}
&=&
\frac{1}{\mathcal
Z_c}\int_{-\pi}^{\pi}\!\frac{\rmd\phi}{2\pi} \rme^{\rmi\phi k_i}\int\!\{\rmd W
\rmd\hat{W}\}~\rme^{\rmi(N-1)\sum_{s k}\int\!\rmd\phi^\prime
\hat{W}_{\!s k}(\phi^\prime) W_{\!s
k}(\phi^\prime)-\bra k\ket\sum_{k}p_{k} Q(k_i,k)+\order(N^{-1})} \nonumber
\\
\hspace*{-25mm} && \times \rme^{\frac{1}{2}(N-2)\bra k\ket \sum_{s^\prime k^\prime
s^\pprime k^\pprime}\int\!\rmd\phi^\prime \rmd\phi^\pprime W_{\!s^\prime
k^\prime}(\phi^\prime) W_{\!s^\pprime k^\pprime}(\phi^\pprime)
Q(k^\prime,k^\pprime)[ \rme^{-\rmi(2\psi s^\prime
s^\pprime+\phi^\prime+\phi^\pprime)}-1 ]} \nonumber
\\
\hspace*{-25mm}&&\times \rme^{-\frac{1}{4}\bra k\ket^2\sum_{s^\prime k^\prime
s^\pprime k^\pprime }\int\!\rmd\phi^\prime d\phi^\pprime
W_{\!s^\prime k^\prime}(\phi^\prime) W_{\!s^\pprime
k^\pprime}(\phi^\pprime) Q^2(k^\prime,k^\pprime)[ \rme^{-\rmi(2\psi
s^\prime s^\pprime+\phi^\prime+\phi^\pprime)}-1]^2} \nonumber
\\
\hspace*{-25mm} &&\times \rme^{-\frac{1}{2}\bra k\ket\sum_{s^\prime
k^\prime}\int\!\rmd\phi^\prime W_{\!s^\prime k^\prime}(\phi^\prime)
Q(k^\prime,k^\prime) [\rme^{-2\rmi(\psi s^{\prime}+\phi^\prime)}-1]
+N\sum_k p_k\log\int_{-\pi}^\pi\!\frac{d\phi^\prime}{2\pi}e^{i[\phi^\prime
 k-\hat{W}_{0 k}(\phi^\prime)]}
}
 \nonumber
\\
\hspace*{-25mm} && \times
\exp\left\{
\bra k\ket\sum_{k} p_{k}Q(k_i,k)\frac{
 \int_{-\pi}^{\pi}\!\rmd\phi^\prime \rme^{\rmi[\phi^\prime
 k-\hat{W}_{1 k}(\phi^\prime)-\phi]}}
 { \int_{-\pi}^{\pi}\!\rmd\phi^\prime \rme^{\rmi[\phi^\prime
 k-\hat{W}_{0 k}(\phi^\prime)]}}-\log\int_{-\pi}^{\pi}\!\frac{\rmd\phi^\prime}{2\pi}\rme^{\rmi[\phi^\prime
 k_i-\hat{W}_{0 k_i}(\phi^\prime)]}
 \right\}
 \nonumber\\
 \hspace*{-25mm}
 &=& \frac{1}{\mathcal
Z_c}\int_{-\pi}^{\pi}\!\frac{\rmd\phi}{2\pi} \rme^{\rmi\phi k_i}\int\!\{\rmd W
\rmd\hat{W}\}~\rme^{N\Psi(W,\hat{W},\psi)+\Phi(W,\hat{W},\psi)+\Omega(\hat{W},k_i,\phi)+\order(N^{-1})}
\end{eqnarray}
with
\begin{eqnarray}
\hspace*{-20mm} \Psi(W,\hat{W},\psi)&=& \rmi\sum_{s k}\int_{-\pi}^\pi\!\rmd\phi~
\hat{W}_{\!s k}(\phi) W_{\!s k}(\phi)+ \sum_{k}p_{k}\log
\int_{-\pi}^{\pi}\!\frac{\rmd\phi}{2\pi} \rme^{\rmi[\phi
 k-\hat{W}_{0 k}(\phi)]}
 \nonumber
\\
\hspace*{-20mm} &&+ \frac{1}{2}\bra k\ket\sum_{s k s^\prime
k^\prime}\int\!\rmd\phi \rmd\phi^\prime~ W_{\!s k}(\phi) W_{\!s^\prime
k^\prime}(\phi^\prime) Q(k,k^\prime)[ \rme^{-\rmi(2\psi s
s^\prime+\phi+\phi^\prime)}\!-\!1 ]
\label{eq:Psi}
\\
\hspace*{-20mm}
 \Phi(W,\hat{W},\psi)&=& -\bra k\ket\sum_{s k s^\prime
k^\prime}\int\!\rmd\phi \rmd\phi^\prime W_{\!s k}(\phi) W_{\!s^\prime
k^\prime}(\phi^\prime) Q(k,k^\prime)[ \rme^{-\rmi(2\psi s
s^\prime+\phi+\phi^\prime)}-1 ] \nonumber
\\
\hspace*{-20mm}&&-\frac{1}{4}\bra k\ket^2\sum_{s k s^\prime k^\prime
}\int\!\rmd\phi \rmd\phi^\prime W_{\!s k}(\phi) W_{\!s^\prime
k^\prime}(\phi^\prime) Q^2(k,k^\prime)[ \rme^{-\rmi(2\psi s
s^\prime+\phi+\phi^\prime)}-1]^2
\\
\hspace*{-20mm} &&-\frac{1}{2}\bra k\ket\sum_{s k}\int\!\rmd\phi~ W_{\!s
k}(\phi) Q(k,k) [\rme^{-2\rmi(\psi s+\phi)}-1] -\rmi\sum_{s
k}\int\!\rmd\phi~ \hat{W}_{\!s k}(\phi) W_{\!s k}(\phi)
\nonumber
\\
\hspace*{-20mm} \Omega(\hat{W},k_i,\phi) &=& \bra k\ket\sum_{k}
p_{k}Q(k_i,k)\Big[\frac{
 \int_{-\pi}^{\pi}\!\rmd\phi^\prime \rme^{\rmi[\phi^\prime
 k-\hat{W}_{1 k}(\phi^\prime)-\phi]}}
 { \int_{-\pi}^{\pi}\!\rmd\phi^\prime \rme^{\rmi[\phi^\prime
 k-\hat{W}_{0 k}(\phi^\prime)]}}
\!-\!1\Big]-\log\int_{-\pi}^{\pi}\!\frac{\rmd\phi^\prime}{2\pi}\rme^{\rmi[\phi^\prime
 k_i-\hat{W}_{0 k_i}(\phi^\prime)]}
 \nonumber
 \\[-2mm]
 \hspace*{-20mm}&&
\end{eqnarray}
Using the normalization identity $\hat{P}_i(0)=1$ we may then also write
\begin{eqnarray}
\hspace*{-15mm}
 \hat{P}_i(\psi)
 &=& \frac{\int_{-\pi}^{\pi}\!\rmd\phi~ \rme^{\rmi\phi k_i}\int\!\{\rmd W
\rmd\hat{W}\}~\rme^{N\Psi(W,\hat{W},\psi)+\Phi(W,\hat{W},\psi)+\Omega(\hat{W},k_i,\phi)+\order(N^{-1})}}{
\int_{-\pi}^{\pi}\!\rmd\phi~ \rme^{\rmi\phi k_i}\int\!\{\rmd W
\rmd\hat{W}\}~\rme^{N\Psi(W,\hat{W},0)+\Phi(W,\hat{W},0)+\Omega(\hat{W},k_i,\phi)+\order(N^{-1})}}
\end{eqnarray}
and, upon defining $P(r|k)=P(k,r)/p_k$:
\begin{eqnarray}
  \hspace*{-15mm}
P(r|k)
 &=& \int_{-\pi}^{\pi}\!\frac{\rmd\psi}{2\pi}\rme^{\rmi\psi r} L_k(\psi)\\
 \hspace*{-15mm}
 L_k(\psi)&=&
 \lim_{N\to\infty} \frac{\int\!\{\rmd W
\rmd\hat{W}\}~\rme^{N\Psi(W,\hat{W},\psi)+\Phi(W,\hat{W},\psi)}\int_{-\pi}^{\pi}\!\rmd\phi~
\rme^{\rmi\phi k+\Omega(\hat{W},k,\phi)}}{\int\!\{\rmd W
\rmd\hat{W}\}~\rme^{N\Psi(W,\hat{W},0)+\Phi(W,\hat{W},0)}
\int_{-\pi}^{\pi}\!\rmd\phi~ \rme^{\rmi\phi k+\Omega(\hat{W},k,\phi)}}
\label{eq:L_k}
\end{eqnarray}
We next need to find the saddle-point(s) of the function (\ref{eq:Psi}), by variation of $\{W,\hat{W}\}$.
Functional differentiation with respect to
 $W$ and $\hat{W}$ gives the following equations, respectively:
\begin{eqnarray}
i \hat{W}_{\!s k}(\phi) &=&
 -\bra k\ket \sum_{s^\prime k^\prime} \int\!\rmd\phi^\prime~ W_{\!s^\prime
k^\prime}(\phi^\prime) Q(k,k^\prime)[ \rme^{-\rmi(2\psi s
s^\prime+\phi+\phi^\prime)}\!-\!1 ]
\\
 W_{\!s k}(\phi) &=& \delta_{s 0} p_{k} \frac{ \rme^{\rmi[\phi
 k-\hat{W}_{0 k}(\phi)]}}
 { \int_{-\pi}^{\pi}\!\rmd\phi^\prime~ \rme^{\rmi[\phi^\prime
 k-\hat{W}_{0 k}(\phi^\prime)]}}
\end{eqnarray}
Upon eliminating $\hat{W}$ and defining $W_{\!s k}(\phi) = \delta_{s 0} p_{k}\chi_k(\phi)$,
we obtain an equation for $\chi_k(\phi)$ only:
\begin{eqnarray}
 \chi_k(\phi)&=& \frac{ \rme^{\rmi\phi
 k+ \bra k\ket\sum_{k^\prime}
p_{k^\prime}
 Q(k,k^\prime)\rme^{-\rmi\phi}\int\! \rmd\phi^\prime~\chi_{k^\prime}(\phi^\prime)
\rme^{-\rmi\phi^\prime}}}
 { \int_{-\pi}^{\pi}\!\rmd\phi^\prime~ \rme^{\rmi\phi^\prime
 k+ \bra k\ket\sum_{k^\prime}
p_{k^\prime}
 Q(k,k^\prime)\rme^{-\rmi\phi^\prime}\int\! \rmd\phi^\pprime~\chi_{k^\prime} (\phi^\pprime)
\rme^{-\rmi\phi^\pprime} }}
\end{eqnarray}
One defines $a_k=\int\!
\rmd\phi^\prime~\chi_k(\phi^\prime) \rme^{-\rmi\phi^\prime}$ and $b_k=
\bra k\ket\sum_{k^\prime} p_{k^\prime}
 Q(k,k^\prime)a_{k^\prime}$, and
finds after some simple manipulations that $a_0=0$ and $a_{k>0}= k/b_k$. This leaves a closed equation for the $b_k$, which
shows that $\lim_{N\to\infty}b_k=\bra k\ket F(k)$, see  (\ref{eq:RS3b}), and a corresponding formula for $\chi_k(\phi)$:
 \begin{eqnarray}
 b_k&=&
\bra k\ket \sum_{k^\prime>0} p_{k^\prime}
 Q(k,k^\prime)k^\prime/b_{k^\prime}
 \\
 \chi_k(\phi)&=& \frac{k!}{2\pi}(b_k\rme^{-\rmi\phi})^{-k}\exp[b_k \rme^{-\rmi\phi}]
\end{eqnarray}
At the relevant saddle-point, we find as a direct consequence of
the form $W_{sk}(\phi)=\delta_{s0}p_k \chi_k(\phi)$ that  the
functions $\Psi(W,\hat{W},\psi)$ and $\Phi(W,\hat{W},\psi)$ are
both independent of the variable $\psi$. This ensures that
expression (\ref{eq:L_k}) is well-defined, but it also gives us
$L_k(\psi)=1$, and hence
\begin{eqnarray}
P(r|k)
 &=& \delta_{r 0}
\label{eq:no_loops}
\end{eqnarray}
We conclude that in our ensemble (\ref{eq:newer_connectivity}) the fraction of nodes in a loop of length
three vanishes in the limit $N\to\infty$, independent of the degree distribution $p(k)$ and independent of
the choice made for the deformation function $Q(k,k^\prime)$.

\end{document}